\documentclass[fleqn,usenatbib]{mnras}

\usepackage{newtxtext,newtxmath}

\usepackage[T1]{fontenc}

\DeclareRobustCommand{\VAN}[3]{#2}
\let\VANthebibliography\thebibliography
\def\thebibliography{\DeclareRobustCommand{\VAN}[3]{##3}\VANthebibliography}

\usepackage{graphicx}	
\usepackage{amsmath}	
\usepackage[noabbrev, nameinlink]{cleveref} 
\Crefname{section}{Sect.}{Sects.}
\Crefname{figure}{Fig.}{Figs.}
\Crefname{equation}{Eq.}{Eqs.}
\crefname{figure}{Figure}{Figures}

\title[NOTT: Characterising hot exozodiacal dust]{Asgard/NOTT: L-band nulling interferometry at the VLTI\\[0.5em] IV. Prospects for characterising hot exozodiacal dust}

\author[K. Ollmann et al.]{
K. Ollmann,$^{1}$\thanks{E-mail: kollmann@astrophysik.uni-kiel.de}
R. Laugier,$^{2}$
J. P. Scott,$^{3}$
T. A. Stuber,$^{3}$
S. Ertel,$^{3,4}$
D. Defrère,$^{2}$
F. Kirchschlager,$^{5,6}$
\newauthor
\ O. Absil,$^{7}$
G. Garreau,$^{8}$
J. C. Augereau,$^{9}$
W. Danchi,$^{10}$
T. D. Pearce,$^{11}$
and S. Wolf$^{1}$
\\ \\
$^{1}$Institute of Theoretical Physics and Astrophysics, University of Kiel, Leibnizstr. 15, 24118 Kiel, Germany\\
$^{2}$Institute of Astronomy, KU Leuven, Celestijnenlaan 200D, 3001 Leuven, Belgium\\
$^{3}$Department of Astronomy and Steward Observatory, The University of Arizona, 933 North Cherry Ave, Tucson, AZ 85721, USA\\
$^{4}$Large Binocular Telescope Observatory, The University of Arizona, 933 North Cherry Ave, Tucson, AZ 85721, USA\\
$^{5}$Sterrenkundig Observatorium, Ghent University, Krijgslaan 281-S9, B-9000 Gent, Belgium\\
$^{6}$Hamburger Sternwarte, University of Hamburg, Gojenbergsweg 112, 21029 Hamburg, Germany\\
$^{7}$STAR Institute, Université de Liège, Allée du Six Août 19c, B-4000 Liège, Belgium\\
$^{8}$Exoplanets and Habitability, ETH Zürich, Wolfgang-Pauli-Str. 27, 8093 Zürich, Switzerland\\
$^{9}$Université Grenoble Alpes, CNRS, IPAG, 38000 Grenoble, France\\
$^{10}$NASA Goddard Space Flight Center, 8800 Greenbelt Road, Greenbelt, MD 20771-2400, USA\\
$^{11}$Department of Physics, University of Warwick, Gibbet Hill Road, Coventry CV4 7AL, UK
}

\date{Accepted 2026 October 22. Received 2026 October 22; in original form 2026 March 12}

\pubyear{2026}

\begin{document}
\label{firstpage}
\pagerange{\pageref{firstpage}--\pageref{lastpage}}
\maketitle

\begin{abstract}
Near-infrared interferometric surveys have revealed extended emission around the photosphere of \(\approx 40\) main-sequence stars, attributed to the presence of hot exozodiacal dust (HEZD). A promising technique for investigating this phenomenon is nulling interferometry. The Very Large Telescope Interferometer (VLTI) nulling interferometer NOTT (Nulling Observations of exoplaneTs and dusT) is designed to characterise exoplanets and HEZD in the \(L'\) band (\(3.5-4.0\,\mu\)m). We aim to assess NOTT’s potential for the detection and characterisation of HEZD with various instrumental and model parameter configurations. We model the HEZD as an optically thin dust ring, using the nearby Sun-like star \(\tau\) Ceti with HEZD as a reference model. Applying the end-to-end nulling tool SCIFYsim to model the instrumental response of NOTT to simulated HEZD brightness distributions, we evaluate how variations in HEZD and instrumental parameters affect NOTT’s key observables: the photon flux rate and the null depth. We demonstrate that NOTT is capable of disentangling stellar and dust contributions to the signal and detecting HEZD. The null depth and photon flux rate are sensitive to the HEZD geometry, with inclination, position angle, and ring location having the strongest impact on the signal. Although the detailed temporal signatures are target-specific, the results can be generalised to other systems. NOTT has the potential to deliver unprecedented sensitivity for HEZD observations, granting access to detailed information through spectral dispersion, beam-combination schemes, and exploitation of sky rotation. This enables more sensitive HEZD detections than is possible with current instruments, and improved spatial constraints on the dust, as well as variability studies.
\end{abstract}



\section{Introduction}

High-precision long-baseline interferometric observations of nearly 200 main-sequence stars, achieving angular resolutions of approximately \(10\) mas, have revealed an interferometric signature in the near-infrared (NIR) photometric \(H\) and \(K\) bands for \(\approx 40\) of these targets (e.g., \citealt{Absil2006, Absil2013, Absil2021}; \citealt{Ertel, Ertel2016}; \citealt{Nunez}). This signature is interpreted as the spatially extended radiation originating from hot exozodiacal dust (HEZD) and translates into an excess over the stellar photospheric emission of at most a few percent. The observed NIR signature has been detected around stars spanning spectral types from G to A and is inferred to originate from regions close to the sublimation radius of the host star (\citealt{DiFolco}).
Although these NIR excesses are commonly explained by submicron-sized grains (\citealt{Kirchschlager2017}), larger particles may also influence the observational appearance of HEZD \citep{Stuber2023}.
Since submicron-sized grains are subject to rapid destruction via sublimation or ejection by radiation pressure (e.g., \citealt{Backman}), the persistence of HEZD around main-sequence stars presents a puzzling problem. To resolve this, models invoke either short-term replenishment through exocometary and planetesimal fragmentation (e.g., \citealt{Bonsor2012, Bonsor2014}; \citealt{Faramaz}; \citealt{Sezestre}) or grain trapping mechanisms facilitated by gas or magnetic fields (e.g., \citealt{Lebreton}; \citealt{Su2013}; \citealt{Rieke}; \citealt{Stamm}; \citealt{Kimura}; \citealt{Pearce2020, Pearce}). Furthermore, silicate grains are unlikely to be a dominant component, as their characteristic mid-infrared (MIR) spectral features would conflict with the faint MIR excess observed (\citealt{Absil2006}; \citealt{Akeson}; \citealt{Kirchschlager2017}; \citealt{Stuber2023}).
Beyond its intrinsic physical properties, HEZD provides critical insights into the broader architecture of planetary systems (\citealt{Kral}). However, its presence also poses a potential challenge for future exo-Earth imaging missions, such as HWO (Habitable Worlds Observatory; \citealt{Ertel2025}) or LIFE (Large Interferometer for Exoplanets; \citealt{Quanz2022}). Specifically, HEZD can induce coronagraphic leakage or parasitic signals, elevate photon noise levels within the habitable zone, and complicate the polarimetric characterisation of close-in exoplanets (\citealt{Ollmann}).\\
\\
Detecting HEZD requires isolating faint circumstellar emission from the dominant stellar flux at small angular separations. Nulling interferometry (\citealt{Bracewell}) addresses this challenge by explicitly suppressing the on-axis light. By adjusting the phases of incoming beams, the technique places a destructive interference fringe (the "null") directly onto the star so that the off-axis radiation from the circumstellar environment is revealed, with its transmission modulated by the specific geometry of the interferometric fringe pattern.
The most productive ground-based nulling interferometers to date have operated in the \(N\) band, including KIN (Keck Interferometer Nuller; \citealt{Serabyn}; \citealt{Colavita}) and LBTI (Large Binocular Telescope Interferometer; \citealt{Hinz}; \citealt{Isbell2024}). These instruments are particularly sensitive to warm habitable-zone dust and have enabled major surveys (KIN: \citealt{Millan-Gabet}; \citealt{Mennesson, Mennesson2014}; LBTI: \citealt{Ertel2020b}) as well as detailed characterisations of individual systems (\citealt{Lebreton}; \citealt{Lebreton2016}; \citealt{Defrere2021}; \citealt{Garreau2025}). Moreover, KIN data have also been used to constrain the properties of HEZD (\citealt{Kirchschlager2017}; \citealt{Stuber2023}).
\\
\\
The NOTT (Nulling Observations of exoplaneTs and dusT; \citealt{Defrere2018, Defrere2022}) is a high-contrast visitor instrument for the Very Large Telescope Interferometer (VLTI; \citealt{Schoeller}; \citealt{Haubois2020}). NOTT employs nulling beam-combination techniques in the \(L'\) band (\(3.5-4.0\,\mu\)m) to characterise exoplanets and HEZD. By combining either the four 8.2\,m UTs (Unit Telescopes) or the four 1.8\,m ATs (Auxiliary Telescopes) over baselines of up to 200\,m, the nulling interferometric approach provides a significant sensitivity improvement for the detection of circumstellar emission compared to FLUOR (Fiber-Linked Unit for Optical Recombination; \citealt{CoudeDuForesto1997}) and PIONIER (Precision Integrated-Optics Near-infrared Imaging ExpeRiment; \citealt{LeBouquin2011}). As part of the Asgard suite (\citealt{Martinod2023}), NOTT relies on the high-precision fringe tracker HEIMDALLR (High-Efficiency Multiaxial Do-it ALL Recombiner; \citealt{Ireland2018}; \citealt{Taras2024}) and the Baldr wavefront sensor (\citealt{Courtney-Barrer}) to reach its target performance. In addition, it operates alongside the beam combiner BIFROST (Beam-combination Instrument for studying the Formation and fundamental paRameters of Stars and planeTary systems; \citealt{Kraus}) and the Solarstein laser alignment calibrator (\citealt{Taras2024}).
In the \(L'\) band regime, the relatively low thermal background from the atmosphere and telescope optics compared to the \(N\) band creates favourable conditions for detecting young giant planets and HEZD via their enhanced contrast against the host star relative to the \(H\) and \(K\) bands. Of the two HEZD studies in the \(L\) band (\citealt{Kirchschlager2020}; \citealt{Ollmann2025}), a clear detection was reported only by \citealt{Kirchschlager2020}, with a dust-to-star flux ratio up to a factor of five higher than typical \(H\) band values, where most detections have been made using constructive long-baseline interferometry. Combined with the tenfold improvement in contrast achieved through nulling interferometry, NOTT \(L'\) band observations are expected to reach an overall sensitivity to HEZD that is approximately 50 times higher than that of previous \(H\) band observations (\citealt{Defrere2022}).\\
\\
The performance of NOTT for detecting exoplanets has already been well-characterised (\citealt{Defrere2024}). The beam combination mode used for these observations, however, is primarily sensitive to asymmetric brightness distributions, whereas the HEZD distribution is expected to exhibit a high degree of symmetry, although its extent is currently unknown. To simplify the modelling, we therefore assume a symmetric HEZD distribution, which reduces the number of free parameters compared to an asymmetric model. To explore the potential of NOTT to detect and characterise HEZD, we introduce, for the first time, alternative beam combination schemes that are sensitive to such symmetric dust geometries. From this analysis, we identify one configuration that combines a strong reduction of unwanted instrumental and atmospheric contributions with sensitivity to symmetric structures, making it the most promising option for HEZD detection. Specifically, we investigate NOTT’s primary observables (the photon flux rate and the corresponding null depth) as a function of the hour angle for selected HEZD and instrumental parameters to identify characteristic geometric signatures of HEZD and assess which model parameters can be constrained through their impact on the temporal modulation of the expected signal.
The paper is structured as follows: In \Cref{model}, we outline the theoretical foundation of NOTT's observables, introduce our stellar and HEZD model and present the assumed observing strategy. In \Cref{results}, we analyse the behaviour of these observables under varying dust and instrumental configurations. We discuss the implications of our results in \Cref{discussion} and summarise our findings in \Cref{Summary}.

\section{Model}\label{model} 
In \Cref{Nulling}, we define NOTT's key observables and discuss various beam combination schemes. In \Cref{HEZD}, we define our star-HEZD model and in \Cref{strategy}, we describe the details of our simulated observations.

\subsection{Nulling interferometry with NOTT}\label{Nulling} 

NOTT employs a four-telescope photonic beam combiner arranged as two Bracewell interferometers, implementing a double-Bracewell scheme \citep{Angel}. This design produces eight outputs in total: four photometric and four interferometric. For the stellar light, two of the interferometric outputs correspond to constructive interference, where the stellar light is located on a bright fringe, while the other two correspond to destructive interference, where the stellar light is located on a dark fringe; the latter are referred to as nulled outputs (e.g., see Fig. 2 in \citealt{Sanny2026}). The four input beams are first separated into photometric and interferometric channels, with the photometric channels routed directly to the four photometric outputs (0, 1, 6, and 7). The interferometric channels are combined pairwise ((0, 1) and (2, 3)) at two first-stage directional couplers. Each first-stage coupler imposes a relative $\pi$ phase shift between its two output channels. One output from each first-stage coupler is routed directly to outputs 2 and 5, yielding two independent pairwise-combined interferometric signals. The remaining two first-stage outputs are fed into a second-stage directional coupler, which again imposes a relative $\pi$ phase shift between its inputs. This final coupler produces outputs 3 and 4, which jointly encode information from all six mutual pairs of telescopes. Depending on the combiner mode, either outputs 2 and 5 or outputs 3 and 4 are the nulled outputs (see \Cref{Fig1}). \\
Variations in the detected observables of NOTT result primarily from instrumental effects and background variations. Instrumental fluctuations include wavefront, tip-tilt, and piston variations driven by residual atmospheric dispersion and jitter, while background variations arise from time-varying thermal loads from the atmosphere and telescope mirrors. By applying specific phase shifts to the individual input beams before they enter the beam combiner, different combiner modes can be realised by changing the relative phases of the four input beams. Three of the possible combiner modes are illustrated in \Cref{Fig1} and described in the following: 

\begin{figure} 
\begin{minipage}{0.50\textwidth} 
\includegraphics[width=\textwidth]{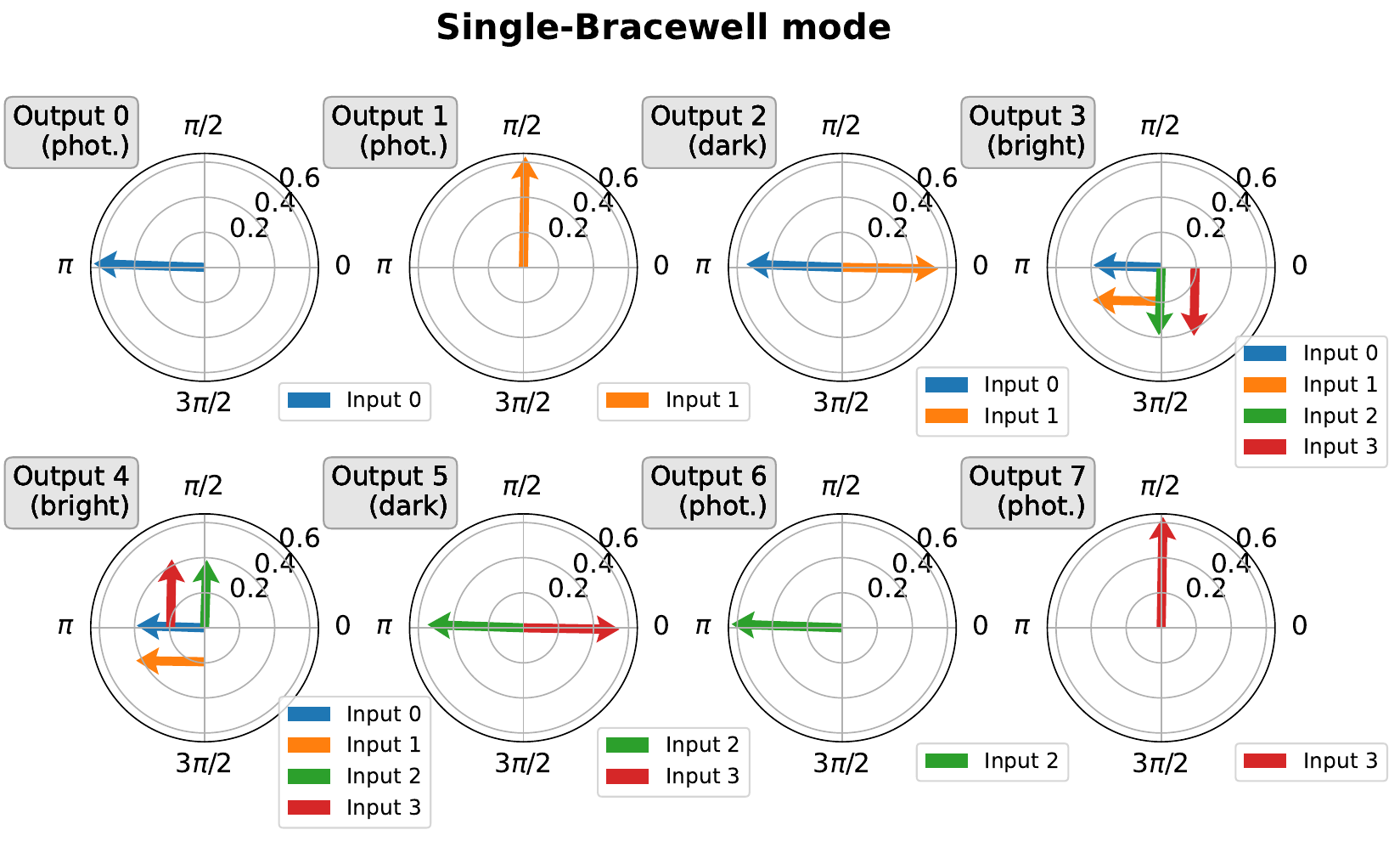} 
\end{minipage} \vspace{0.3cm} 
\begin{minipage}{0.50\textwidth} 
\includegraphics[width=\textwidth]{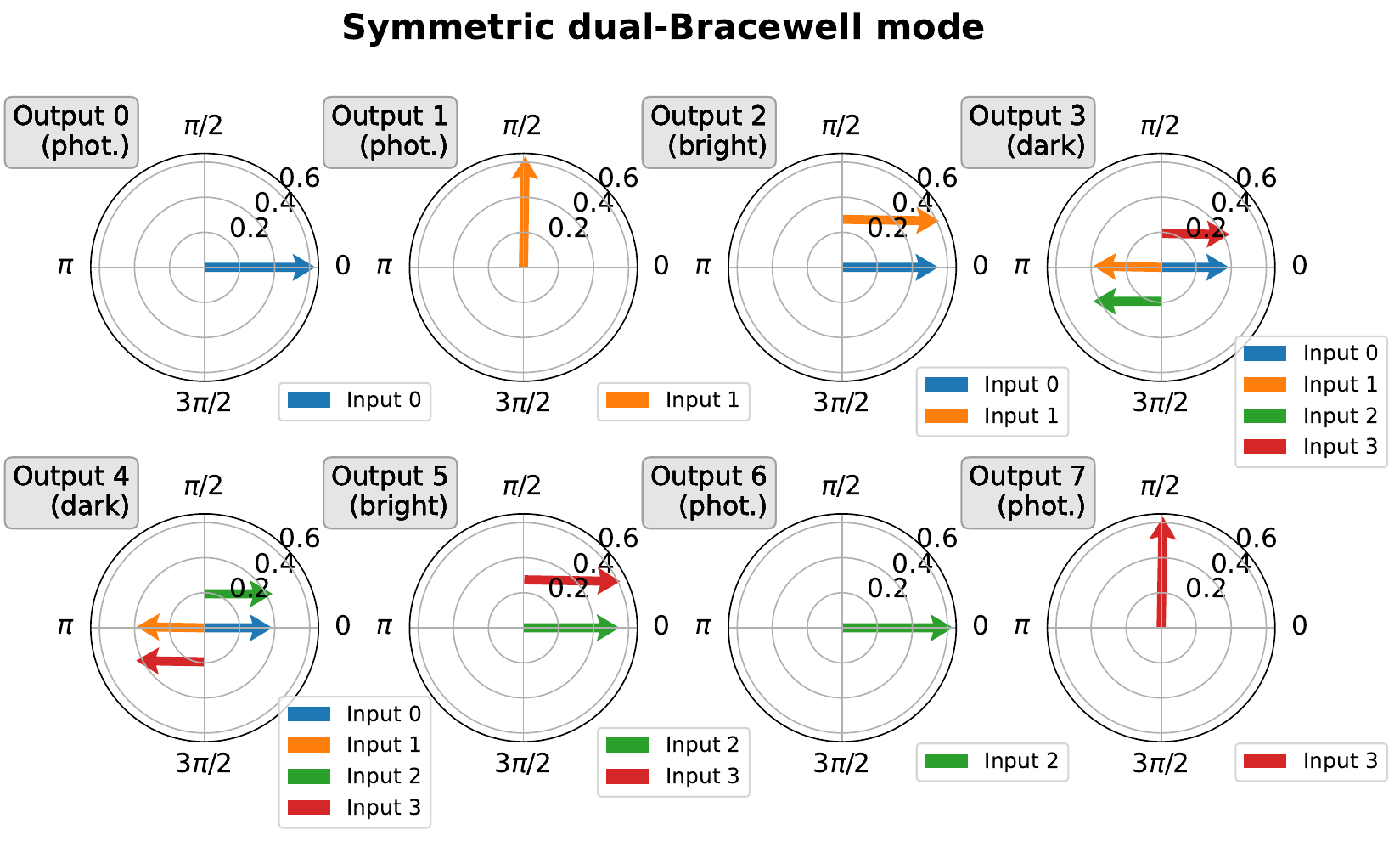} 
\end{minipage} \vspace{0.3cm} 
\begin{minipage}{0.50\textwidth} 
\includegraphics[width=\textwidth]{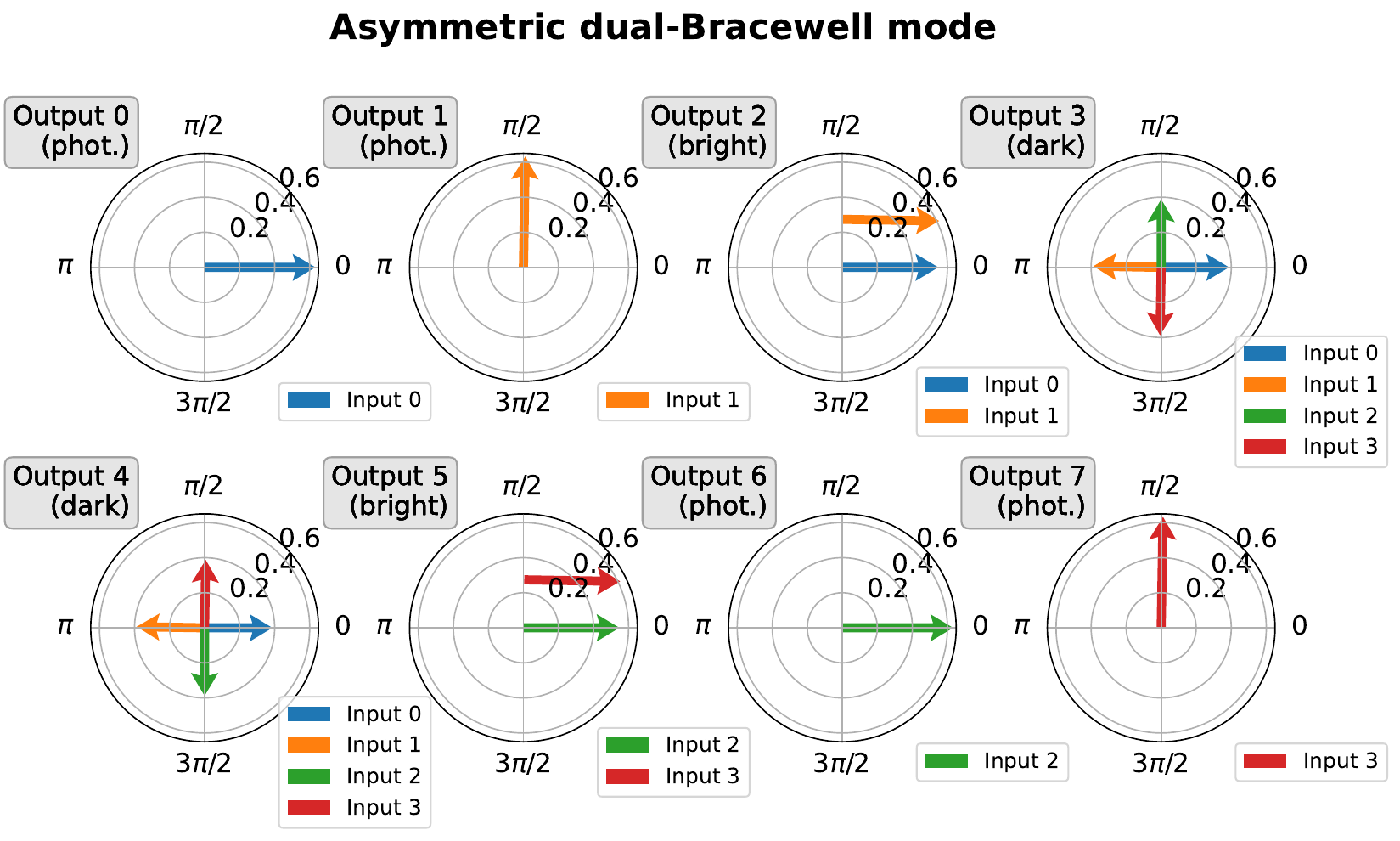} 
\end{minipage}
\caption{Complex matrix diagrams for the three different four-input combiner modes at a wavelength of \(3.75\) \(\mu\)m. Top: Single-Bracewell mode. Mid: Symmetric dual-Bracewell mode. Bottom: Asymmetric dual-Bracewell mode. Each arrow represents the transmission of light from an input to the shown output, with its direction indicating the relative phase orientation of the beams at the outputs of the beam combiner, and its length indicating the relative intensity. The relative phase of \(\pi/2\) between the photometric outputs 0 and 1, as well as between outputs 6 and 7, results from phase offsets applied to the input beams before the beam combiner, which compensate the $\pi/2$ phase shift imposed by the directional couplers on the cross-coupled light, and does not affect the measured intensities. The vector sum of all arrows at a given output determines the resulting interference of the transmitted light. For the stellar light, this interference is constructive at the bright outputs and destructive at the nulled (dark) outputs. Overlapping arrows are displayed side by side for clarity (see \Cref{Nulling} for details).} 
\label{Fig1} 
\end{figure} 

\begin{figure}

    \begin{minipage}{0.52\textwidth}
        \includegraphics[width=\textwidth]{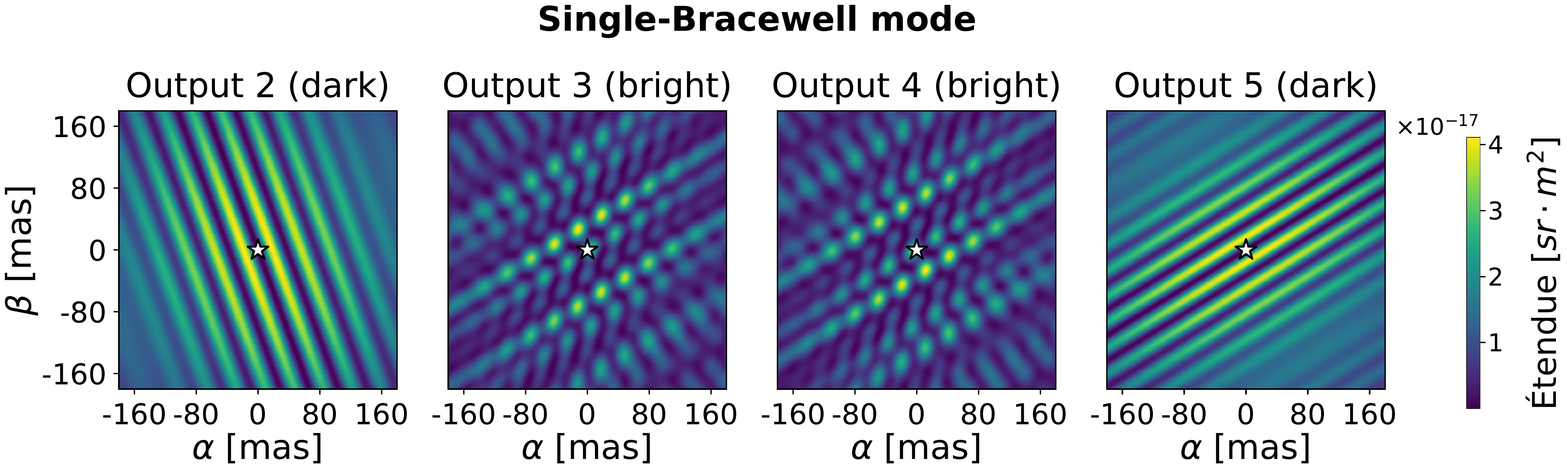}
    \end{minipage}

    \vspace{0.45cm}

    \begin{minipage}{0.52\textwidth}
        \includegraphics[width=\textwidth]{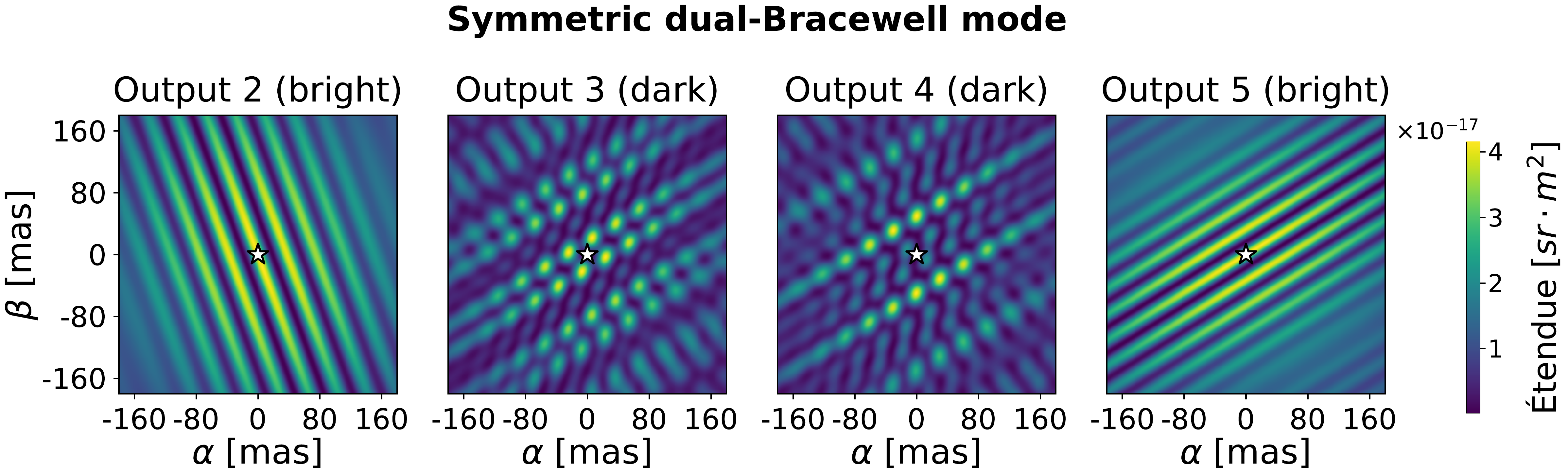}
    \end{minipage}

    \vspace{0.45cm}

    \begin{minipage}{0.52\textwidth}
        \includegraphics[width=\textwidth]{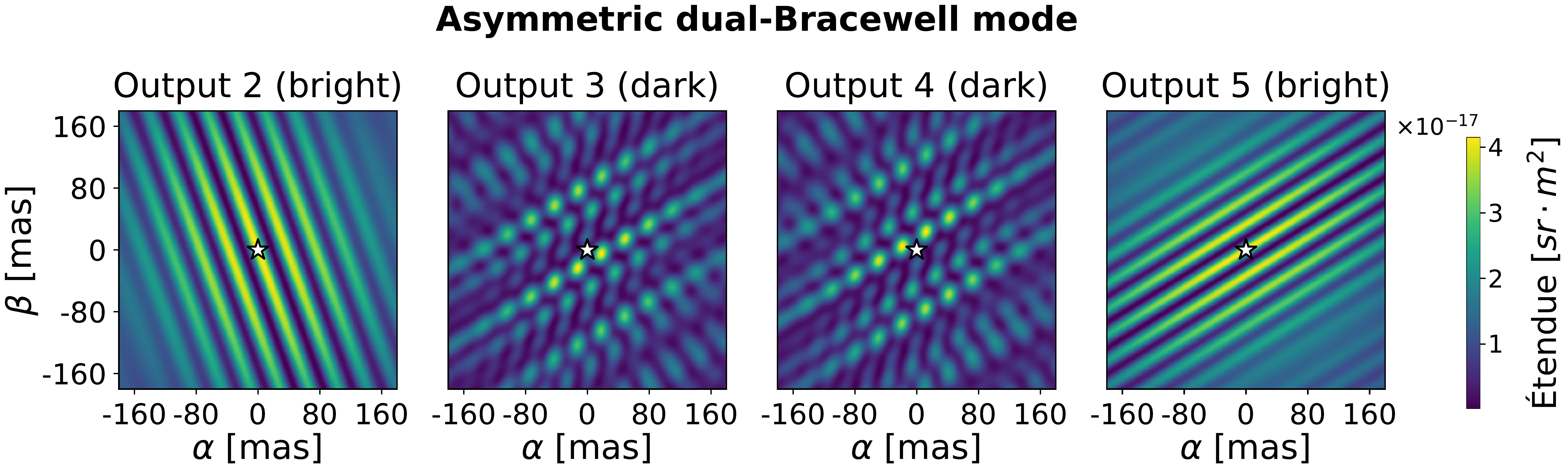}
    \end{minipage}

    \caption{Schematic representation of the transmission maps for the bright and dark outputs with the star located in the central pixel for the AT small configuration, whose six static baselines range from 11 to 34 m. Top: Single-Bracewell mode. Mid: Symmetric dual-Bracewell mode. Bottom: Asymmetric dual-Bracewell mode. All panels are computed monochromatically at a wavelength of \(3.75\ \mu\)m. The values account for the transmission of the individual outputs and are expressed in units of \textit{\'Etendue} (\(\mathrm{sr \cdot m^{2}}\)), where \(\mathrm{sr}\) denotes steradians and \(\mathrm{m^{2}}\) denotes square metres, and can thus be interpreted as the effective collecting capability associated with each individual pixel. The overall decline in \textit{\'Etendue}\ away from the map centre reflects the diffraction-limited FOV of the individual AT apertures, which modulates the fringe pattern produced by the multi-aperture interference.}  \label{FigA1}
\end{figure}

\begin{enumerate} 
\item The \textbf{single-Bracewell mode} (top plot) represents the simplest configuration, in which the four input beams are divided into two pairs, with each pair forming an independent two-telescope Bracewell interferometer. The corresponding dark outputs (2 and 5) are treated independently, providing two simultaneous single-baseline measurements. The relative phase shift between the two beams at each dark output is \(\pi\). This configuration provides high throughput for off-axis emission and preserves spatial information, but offers no inherent mitigation of instrumental phase and amplitude fluctuations and is therefore expected to provide the lowest null-depth precision. 
\item The \textbf{symmetric dual-Bracewell mode} (mid plot) combines the two single-Bracewell nulls into a single configuration, with a relative phase shift of \(\pi\) between the two single-Bracewell nulls before the second-stage directional coupler. Both nulled outputs (3 and 4) contain contributions from all four input beams and therefore exhibit partially correlated instrumental phase and amplitude fluctuations. When the two nulled outputs are subtracted in post-processing, these correlated variations partially cancel, providing a modest reduction of these instrumental and background contributions \citep{Laugier2020,Scott2026}. The resulting transmission pattern retains sensitivity to symmetric and extended circumstellar emission, such as a point-symmetric dust ring, although the combination of multiple baselines reduces the spatial information.

\item The \textbf{asymmetric dual-Bracewell mode} (bottom plot) is realised by tuning the input beam phases so that the two single-Bracewell nulls are routed into the second-stage directional coupler, resulting in two complementary dark outputs (3 and 4). Each dark output contains contributions from all four input beams and all six possible baselines, and these two dark outputs are complex conjugates with partially correlated instrumental fluctuations. Subtracting these outputs (differential asymmetric dual-Bracewell mode, `\(-\)') therefore enables strong cancellation of correlated instrumental and background variations, providing the most robust mitigation of these systematic effects. However, the resulting asymmetric transmission pattern strongly suppresses symmetric and extended emission, making this mode particularly suited to point-like or asymmetric sources such as exoplanets or clumpy dust structures. Adding the two nulled outputs instead (additive asymmetric dual-Bracewell mode, `\(+\)') retains sensitivity to symmetric emission but provides weaker mitigation of these fluctuations.

\end{enumerate} In the following, we introduce the quantities essential for interpreting NOTT's primary observables, while a more detailed description of the mathematical framework underlying the applied nulling formalism is provided in the appendix (\Cref{app:math_details}).\\
Considering the eight outputs of NOTT, the transmission map \(T_m\) for the output \(m\) (\(m = 0, 1, \dots, 7\)) acts as a spatial filter, indicating how flux from different angular positions in the field of view (FOV) is transmitted or nulled (see \Cref{FigA1}). The FOV has a diameter of 430 mas for the ATs at a wavelength of \(3.75\ \mu\)m (\citealt{Martinod2023}). To calculate these transmission maps, we use the end-to-end nulling simulator SCIFYsim (\citealt{Laugier2023}). We note that for the present parameter study, we assume ideal observing conditions and do not inject any instrumental or atmospheric errors into the simulations, allowing us to isolate the geometric and astrophysical signatures of the HEZD. The differential map \(T_{\rm diff}\) (subtraction of the transmission maps for the outputs 3 and 4, see Eq.~\ref{Gl6} in appendix~\ref{app:math_details} for the explicit expression) corresponds to the self-calibrated Kernel null, which provides the aforementioned resistance to instrumental and atmospheric phase errors (\citealt{Martinache}; \citealt{Laugier2020}; \citealt{Chingaipe2023}) and removes the thermal background. One of NOTT's key observables, the photon flux rate collected in output \(m\) (\citealt{Dannert}), is computed as
\begin{equation} \label{Gl7}
I_m(\lambda) =
\sum_{i=0}^{N_\alpha-1}
\sum_{j=0}^{N_\beta-1}
T_m(\alpha_i,\beta_j,\lambda)
J(\alpha_i,\beta_j,\lambda),
\end{equation}
where \(J(\alpha_i,\beta_j,\lambda)\) denotes the surface brightness distribution of the source, in our case a star surrounded by HEZD, evaluated at the angular offset \((\alpha_i,\beta_j)\) from the optical axis. Here, \(i\) and \(j\) denote the pixel indices of the discretised angular coordinates \(\alpha_i\) and \(\beta_j\), with \(N_\alpha\) and \(N_\beta\) denoting the number of pixels along the respective angular coordinates, expressed in mas. The wavelength is denoted by \(\lambda\). The differential photon flux rate is defined as
\begin{align}\label{Gl8}
P_{\mathrm{diff}}(\lambda)
&= \sum_{i=0}^{N_\alpha-1} \sum_{j=0}^{N_\beta-1}
   T_{\mathrm{diff}}(\alpha_i,\beta_j,\lambda)\,
   J(\alpha_i,\beta_j,\lambda)\\
&= \sum_{i=0}^{N_\alpha-1} \sum_{j=0}^{N_\beta-1}
   P(\alpha_i,\beta_j,\lambda). 
\end{align}
This quantity is expressed as the number of detected photons per unit time, \(n_{\mathrm{ph}}/t\), where \(n_{\mathrm{ph}}\) denotes the number of detected photons and \(t\) the time. The photon flux rate is computed by propagating the input photon rate through the instrument, accounting for the transmission losses at each stage of the optical path. It also accounts for the finite FOV of the telescopes and the photonic inputs, such that the summation over $(\alpha_{i},\beta_{j})$ is truncated to the finite FOV set by the aperture. To account for chromatic effects, the input phases are computationally tuned to optimise the suppression of the stellar light across the considered wavelength range (e.g., \citealt{Sharma2020}), which directly propagates into the resulting transmission maps. We refer the reader to \citet{Laugier2023} for a detailed and quantitative analysis of how these chromatic effects impact the overall capabilities of the instrument.
The photon flux rate can be used to compare the detected signal between different observing scenarios, but relating it to calibrated observational quantities requires a suitable reference measurement, such as the stellar photon flux measured in a constructive or photometric output. For a fixed wavelength \(\lambda\), the quantity \(P(\alpha, \beta, \lambda)\) denotes a two-dimensional map of the photon flux density. Finally, the corresponding null depths for the different combiner modes are defined via the photon flux rates as 

\begin{equation} \label{Gl9} 
N_{0,\ \mathrm{SB}}=\frac{b}{1-b}\frac{I_{2,\ \mathrm{SB}}}{I_{0,\ \mathrm{SB}}+I_{1,\ \mathrm{SB}}},
\end{equation}

\begin{equation} \label{Gl10} 
N_{1,\ \mathrm{SB}}=\frac{b}{1-b}\frac{I_{5,\ \mathrm{SB}}}{I_{6,\ \mathrm{SB}}+I_{7,\ \mathrm{SB}}},
\end{equation}

\begin{equation} \label{Gl11} 
N_{\mathrm{sym}}=\frac{b}{1-b}\left(\frac{I_{3,\ \mathrm{sym}}}{I_{0,\ \mathrm{sym}}+I_{1,\ \mathrm{sym}}}-\frac{I_{4,\ \mathrm{sym}}}{I_{6,\ \mathrm{sym}}+I_{7,\ \mathrm{sym}}}\right),
\end{equation}

\begin{equation} \label{Gl12} 
N_{\mathrm{asym}}^{\pm}=\frac{b}{1-b}\left(\frac{I_{3,\ \mathrm{asym}}}{I_{0,\ \mathrm{asym}}+I_{1,\ \mathrm{asym}}}\pm\frac{I_{4,\ \mathrm{asym}}}{I_{6,\ \mathrm{asym}}+I_{7,\ \mathrm{asym}}}\right).
\end{equation}
Here, \(b\) denotes the photometric tap ratio, i.e., the fraction of the input signal directed to the four photometric outputs. The subscripts indicate the combiner mode: "SB" stands for single-Bracewell (each of the null depths \(N_{0,\ \mathrm{SB}}\) and \(N_{1,\ \mathrm{SB}}\) corresponds to a fixed baseline), "sym" for symmetric dual-Bracewell, and "asym" for asymmetric dual-Bracewell. The null depth of a selected observing mode is thus physically determined by the input phase configuration of the beam combiner, which leads to distinct transmission maps and, consequently, to different photon flux rates. The \(+\) and \(-\) signs in \(N_{\mathrm{asym}}^{\pm}\) arise solely from post-processing choices when combining the two nulled outputs of the same asymmetric combiner mode.
Moreover, the null depths defined in this section differ from conventional metrics used by legacy instruments like KIN or LBTI. Instead of normalising by a constructive output, NOTT uses the simultaneous photometric channels (outputs 0, 1, 6, 7). Scaled by the factor \(\frac{b}{1-b}\), these serve as a robust proxy for the total constructive flux, making the single-Bracewell combiner null depth quantitatively consistent with the standard definition. For the dual-Bracewell modes, we define a differential null depth. This differential formulation naturally allows for negative values, where the resulting sign provides valuable geometric information by indicating which spatial regions of the transmission map dominate the astrophysical signal. Unless otherwise stated, the symmetric dual-Bracewell mode is used as the default observational combiner mode for this presentation such that we set \(P_{\mathrm{sym}} = P_{\mathrm{diff}}\).

\subsection{Star and HEZD model} \label{HEZD} 
To evaluate the feasibility of detecting HEZD with NOTT in a realistic scenario, we adopt the nearby (3.65\,pc; \citealt{Gaia}) Sun-like G8V star \(\tau\) Ceti (HD 10700) as a benchmark star. Observations of \(\tau\) Ceti with FLUOR revealed a significant NIR excess of \(1\%\) in the \(K\) band, providing clear evidence for the presence of HEZD \citep{DiFolco}.  
 Owing to its similarity to the Sun and its potential to host terrestrial planets in the habitable zone \citep{Tuomi2013,Feng}, \(\tau\) Ceti is also considered as a key target for future missions such as HWO (\citealt{Tuchow2025}). These characteristics make \(\tau\) Ceti a suitable test case for NOTT observations. The star is modelled as a uniform disc with the radius of \(\tau\) Ceti (\(0.98\)\,mas; \citealt{Korolik2023}). This approach accounts for finite stellar leakage in the nulling interferometric response: as the stellar angular diameter becomes non-negligible relative to the fringe spacing set by the baseline, the central dark fringe no longer fully suppresses the stellar flux, resulting in residual leakage.\\
\\
For the simulation of the HEZD, we adopt a geometrically and optically thin ring model following previous HEZD studies (e.g., \citealt{Kirchschlager2017}). We calculate its surface brightness distribution \(J(\alpha,\beta,\lambda)\) with the simulation tool Debris disks around Main-sequence Stars (DMS; \citealt{DMS}). The dust ring is characterised by an inner radius $R_{\mathrm{in}}$ and an outer radius $R_{\mathrm{out}}$, an inclination $i$ (with $i=0^\circ$ denoting a face-on orientation and  $i=90^\circ$ denoting an edge-on orientation), and a position angle PA that defines the orientation of the semi-major axis of the projected ellipse, measured anticlockwise from East ($0^\circ$). The radial number density between the inner and outer ring radii follows a power-law distribution $n(r) \propto r^{-2}$ with \(r\) as the radial distance between the position of the dust grain and the star. We consider two dust compositions in this study: amorphous carbon \citep{Rouleau} and graphite with the standard 1/3--2/3 approximation \citep{Draine}, corresponding to an orientation-averaged treatment of the anisotropic dielectric properties of randomly oriented graphite grains. This choice follows previous HEZD modelling: silicates are disfavoured owing to their
lower sublimation temperature, whereas graphite and amorphous carbon can persist closer to the star; \citealt{DiFolco} specifically find graphite to provide a better fit than silicate for the HEZD around $\tau$ Ceti. The dust grains are assumed to be compact spheres with a bulk density of $\rho = 2.24\,\mathrm{g\,cm^{-3}}$ for both materials. Their optical properties are computed using the tool miex \citep{Wolf}.\\
Our reference HEZD model is guided by the best-fitting parameter values inferred from the analysis of the $K$ and \(N\) band observations of \(\tau\) Ceti by \citet{Kirchschlager2017}. Specifically, we adopt a ring ranging from $R_{\mathrm{in}} = 0.054\,\mathrm{au}$ to $R_{\mathrm{out}} = 0.06\,\mathrm{au}$, with an inclination of $i = 45^\circ$, a position angle of $\mathrm{PA} = 0^\circ$ and a grain  radius of $a = 0.10\,\mu\mathrm{m}$. At these small distances from the star, the HEZD reaches temperatures close to the sublimation limit of graphite (2000 K, \citealt{Kobayashi}), with $T_{\mathrm{dust}} \approx 1780\,\mathrm{K}$. The HEZD mass of $M_{\mathrm{dust}} = 4.10 \times 10^{-11}\,\mathrm{M_{\oplus}}$ is chosen such that it results in a dust-to-star flux ratio of $1\%$ at an observing wavelength of $3.75\,\mu\mathrm{m}$. For all other parameter configurations explored in this study for $\tau$ Ceti, we therefore scale the HEZD mass to maintain this dust-to-star flux ratio. While this scaling does not reproduce the observed $K$ band excess, it allows us to consistently compare the characteristic HEZD signatures across the different parameter configurations without affecting the characteristics of NOTT's observables for the different telescope and HEZD parameter configurations. An overview of the adopted stellar, HEZD, and instrumental parameters is provided in \Cref{table1}.

\begin{table}
\caption{Parameters of the reference model consisting of the central star \(\tau\) Ceti, the circumstellar HEZD distribution, and the  instrumental configuration of NOTT. The astrometric data for \(\tau\) Ceti were obtained from Gaia DR3 via the ESA Gaia Archive (\citealt{Gaia}), using coordinates (J2016.0, ICRS) with proper-motion  corrections to the observation date (2026 October 02). In the instrument parameters, ``Small'' refers to the compact baseline configuration of the ATs. The telescope input order maps the instrument's four input channels (0, 3, 1, 2) to the VLTI AT station pads (A0, C1, B2, D0), where the channel-to-station mapping is 0 = A0, 1 = B2, 2 = D0, and 3 = C1.}
\label{table1}

\begin{tabular}{l l l}
\hline
 & Parameter & Value \\
\hline
\textbf{\(\tau\) Ceti} & Radius \(R_{\star}\) [mas] & 0.98 \\
 & Effective temperature \(T_{\star}\) [K] & 5344 \\
 & Right ascension [\(^\circ\)] & 26.01 \\
 & Declination [\(^\circ\)] & -15.93 \\
 & Meridian transit time [LST] & 01:44 \\
 & Distance to observer \(d_{\rm obs}\) [pc] & 3.65 \\
\noalign{\vskip 3pt}
\hline
\textbf{HEZD} & Orbital inclination \(i\) [\(^\circ\)] & 45 \\
 & Grain radius \(a\) [\(\mu\)m] & 0.10 \\
 & Dust type & Graphite \\
 & HEZD mass \(M_{\rm dust}\) [\(\mathrm{M_\oplus}\)] & \(4.1\times10^{-11}\) \\
 & Inner ring radius \(R_{\rm in}\) [au] & 0.054 \\
 & Outer ring radius \(R_{\rm out}\) [au] & 0.06 \\
 & Temperature \(T_{\rm dust}\) [K] & 1780 \\
 & Position angle PA [\(^\circ\)] & 0 \\
\hline
\textbf{Instrument} & Telescopes & ATs \\
 & Configuration & Small \\
 & Telescope input order & 0, 3, 1, 2 (A0, C1, B2, D0) \\
 & Combiner mode & Symmetric dual-Bracewell \\
 & Number of observations & 7 \\
 & Photometric tap ratio \(b\) & 0.4 \\
 & Observing wavelength \(\lambda\) [\(\mu\)m] & 3.75\\
\hline
\end{tabular}
\end{table}

\subsection{Observing strategy}\label{strategy}
The simulated NOTT observing sequence of \(\tau\) Ceti spans a total duration of  $7.5\,\mathrm{h}$. It consists of seven standard acquisition snapshots, defined by  repeated CAL-SCI-CAL-SCI-CAL sequences, where SCI denotes an observation of the science target while CAL denotes an observation of a calibrator star. While a single isolated sequence is assumed to last $75\,\mathrm{min}$, the total duration is constrained to $7.5\,\mathrm{h}$ by sharing calibrators between consecutive sequences. The sequence is centred on the meridian transit of the target. Consequently, the hour angle varies from $-3.75\,\mathrm{h}$ to \(+3.75\,\mathrm{h}\). Observations centred on meridian transit are particularly favourable for nulling interferometry. At transit, the target reaches its maximum elevation, minimising the airmass and, consequently, the impact of atmospheric turbulence and thermal background emission. This observing strategy is therefore expected to provide more favourable conditions for phase stability and piston control. Nevertheless, atmospheric and instrumental effects can still affect the achievable precision of the calibrated null depth under real observing conditions. These effects are intentionally not included in the simulations presented here, which are designed to isolate the intrinsic response of the NOTT observables to the HEZD signal and to investigate their potential for constraining the HEZD geometry. The reference model employs the AT small configuration (stations A0, B2, C1, D0), yielding six baselines ranging from 11 to 34\,m, of which the pairs \(\mathrm{A}0-\mathrm{C}1\) and \(\mathrm{B}2-\mathrm{D}0\) form the primary single-Bracewell null baselines for the chosen telescope input order used in this study (see \Cref{table1}).

\section{Results}\label{results}
In this section, we present the results of our parameter study, evaluating the observables of NOTT under various instrumental conditions and for different HEZD parameters. In \Cref{referenceresults}, we discuss the impact of the reference model parameters for \(\tau\) Ceti with HEZD on the resulting null depth and photon flux rate. Subsequently, in \Cref{telescope}, we examine the photon flux rate and null depth for selected instrumental configurations and observing wavelengths, while in \Cref{hezd parameters}, we explore the same observables for different geometrical HEZD parameter values. Throughout this analysis, any parameter not explicitly varied adopts the value of the corresponding reference model parameter. In \Cref{stars}, we compare the null depth and photon flux rate for the systems \(\kappa\) Tuc A  and Fomalhaut with those obtained for \(\tau\) Ceti. Finally, in \Cref{Kappatuc}, we analyse the expected null depth for HEZD around \(\kappa\) Tuc A using parameter values compatible with previous \(L\) band measurements. The stellar contribution, including the associated stellar leakage, is assumed to be removed through calibration, leaving only the contribution from the HEZD in the calibrated signal, with a calibration performance comparable to that achieved in previous nulling interferometric observations. Thus, unless explicitly stated otherwise, the results presented in this section refer to the HEZD contribution to the null depth and photon flux rate. As stated in \Cref{model}, the resulting sensitivity represents the potential performance under idealised conditions.

\subsection{Reference model}\label{referenceresults}

\begin{figure}
\centering
\includegraphics[width=\columnwidth]{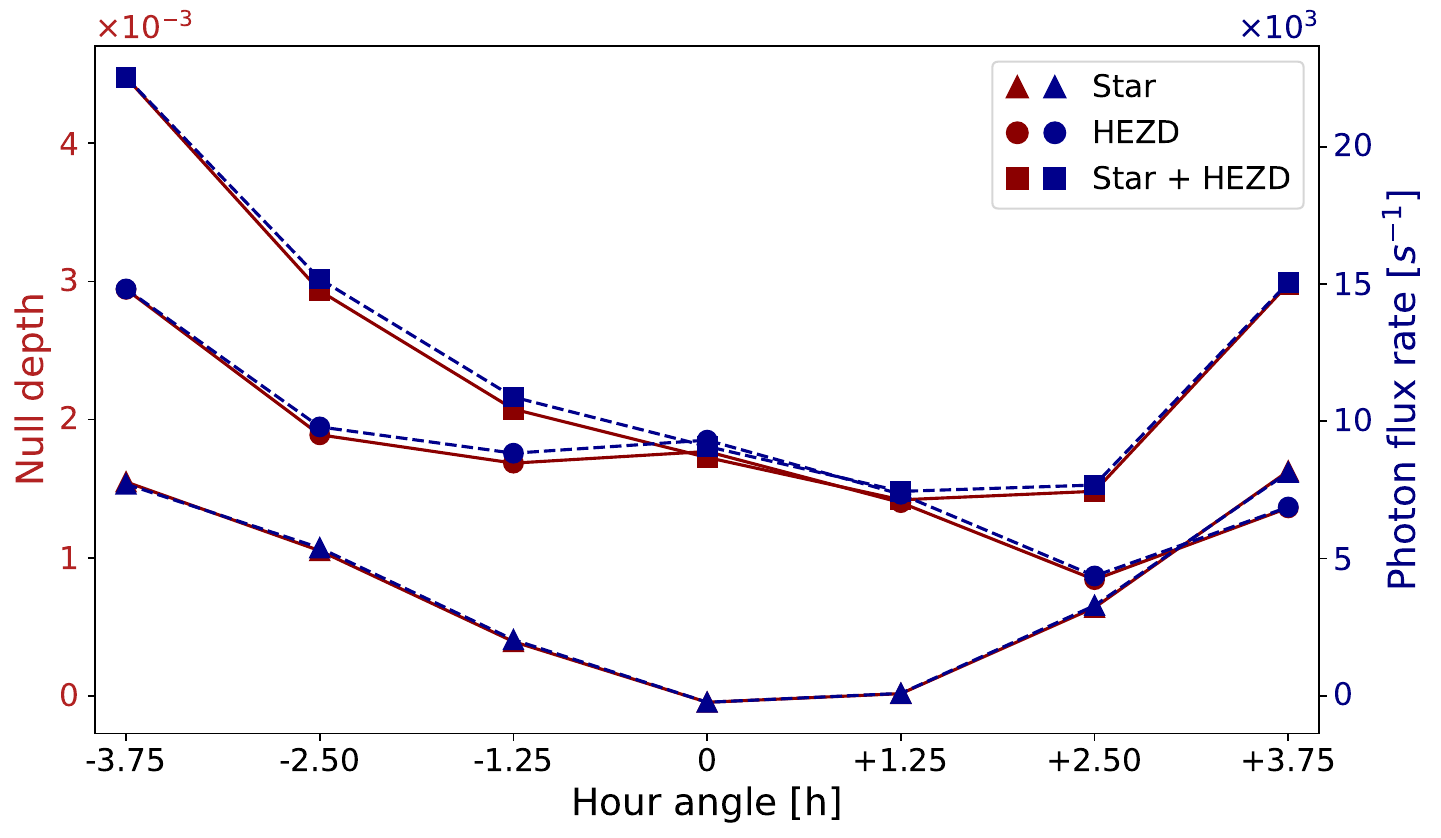}

\vspace{0.3cm}

\includegraphics[width=1.01\columnwidth]{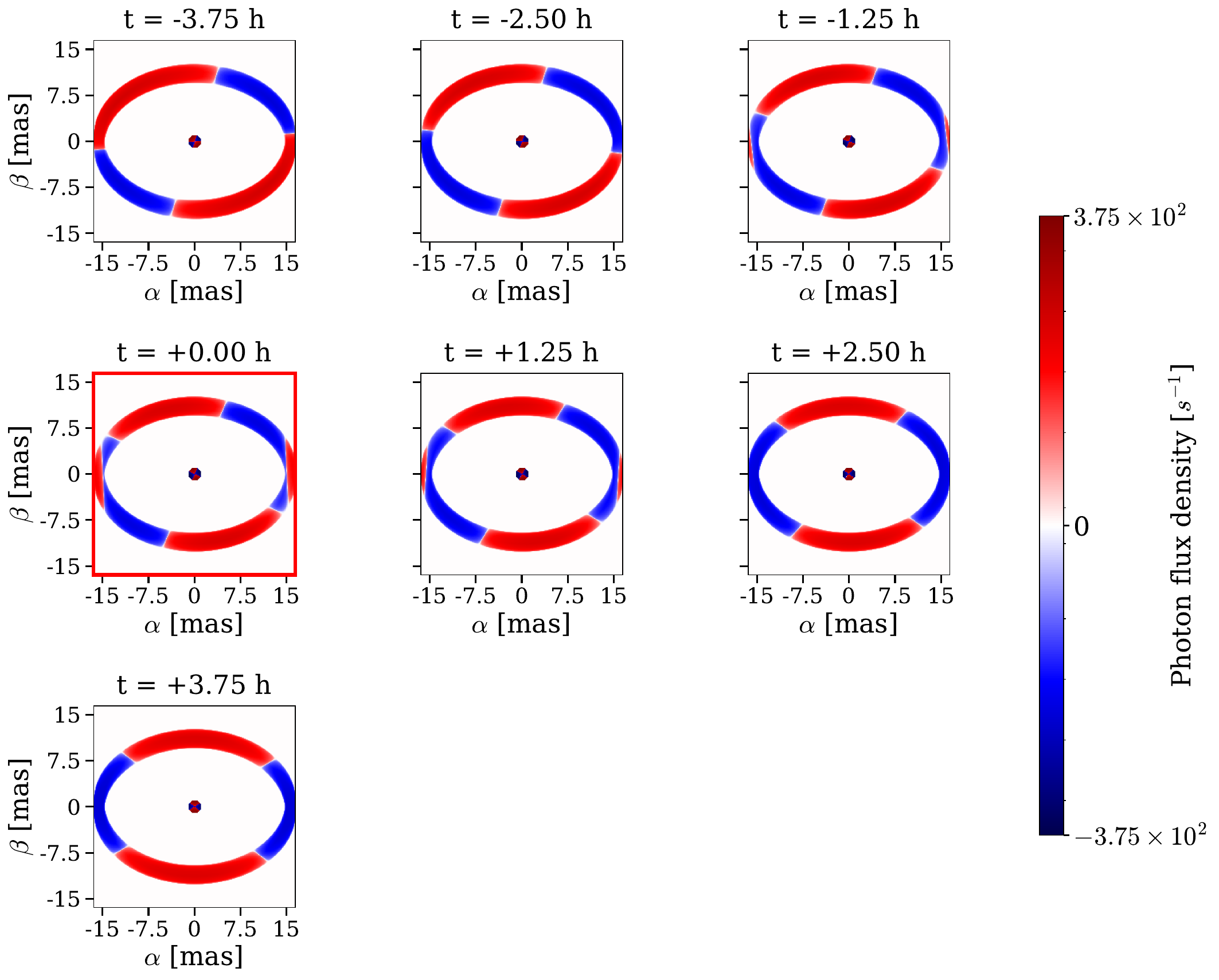}

\caption{Top: Null depth \(N_{\mathrm{sym}}\) and photon flux rate \(P_{\mathrm{sym}}\) for \(\tau\) Ceti with HEZD and the reference parameter values from \Cref{table1} as a function of the hour angle. The temporal modulations of both quantities track each other closely because the null depth is derived by scaling the differential photon flux rate with the stellar flux measured in the photometric outputs. Bottom: Photon flux density \(P\) for \(\tau\) Ceti with HEZD shown for seven snapshots at equal time intervals. The snapshot corresponding to the meridian transit is marked by an outlining box (see \Cref{referenceresults} for details).}
\label{Fig2}
\end{figure}

The top plot of \Cref{Fig2} illustrates the temporal evolution of the photon flux rate (\hyperref[Gl8]{Eq.~\ref*{Gl8}}) and the null depth (\hyperref[Gl11]{Eq.~\ref*{Gl11}}) for the reference HEZD model and instrumental configuration (see \Cref{table1}). Both observables show a clear temporal correlation, driven by a modulation pattern induced by the time-dependent rotation of the instrumental transmission map as the projected baselines change during the observation. For the combined star-HEZD signal, the null depth reaches a maximum value of $\approx 4.4 \times 10^{-3}$ at the beginning of the observation ($t = -3.75\,\mathrm{h}$) and subsequently decreases to a minimum of $\approx 1.5 \times 10^{-3}$ at $t = +1.25\,\mathrm{h}$. The photon flux rate follows an analogous behaviour, varying between a maximum of $\approx 22 \times 10^{3}\, s^{-1}$ and a minimum of $\approx 7 \times 10^{3}\,s^{-1}$. A direct comparison between the star-only and the star–HEZD signals reveals a systematic offset in both observables, which directly traces the additional thermal emission contributed by the HEZD. This isolated HEZD contribution reaches a maximum null depth of $\approx 2.9 \times 10^{-3}$ and a maximum photon flux rate of $\approx 14 \times 10^{3}\,s^{-1}$. With an estimated \(1\sigma\) null depth precision of \(\approx 4 \times 10^{-4}\) for stars with an \(L'\) band magnitude comparable to that of \(\tau\) Ceti in the symmetric dual-Bracewell mode (\citealt{Scott2026}), the signals of the HEZD are well above this value. As demonstrated by \citet{Scott2026}, this precision estimate already accounts for realistic levels of atmospheric piston, residual adaptive optics corrections, and the modelled chromaticity of the NOTT optics. While this value represents the precision of a single measurement and does not explicitly include calibration uncertainties, standard calibration against a reference star is expected to scale these uncertainties by a factor of \(\sqrt{2}\). Even with this adjustment, the signals of the HEZD remain well above the precision threshold, ensuring a robust measurement of the HEZD signature. Notably, around meridian transit, the signal is dominated by the HEZD emission rather than stellar leakage.\\ 
\\
The bottom plot of \Cref{Fig2} shows the  photon flux density maps for the corresponding seven time steps. These maps visualise the interplay between the static geometry of the HEZD and the time-dependent instrumental transmission pattern. Positive values indicate constructive transmission lobes of output 3, while negative values correspond to those of output 4. The fringe patterns appear broader when the projected baselines are short and become progressively narrower for longer baselines; as the target moves away from the meridian transit, the projected baselines shorten again, causing the transmission patterns to widen. At the beginning and end of the observing sequence ($t = \pm 3.75\,\mathrm{h}$), the geometric overlap between the transmission pattern and the ring leads to extended regions of net positive signal, corresponding to the maxima observed in the photon flux rates in  the top plot of \Cref{Fig2}. Finally, we emphasise that, while photon flux density maps are useful for visualising the effects of the time-dependent geometric transmission pattern, NOTT measures only the spatially integrated contribution across the full FOV (the photon flux rate).

\subsection{Instrumental parameters} \label{telescope}
We examine the influence of the instrumental setup on the photon flux rate and the null depth, beginning with the choice of the NOTT combiner modes (\Cref{combinermodes}) and the observing wavelength (\Cref{wavelength}). Subsequently, in \Cref{configuration} and \Cref{input}, we investigate how variations in the telescope configurations and input orders modify the projected transmission maps and consequently the resulting NOTT observables.

\subsubsection{Combiner modes}\label{combinermodes}

\begin{figure}
\includegraphics[scale=0.35]{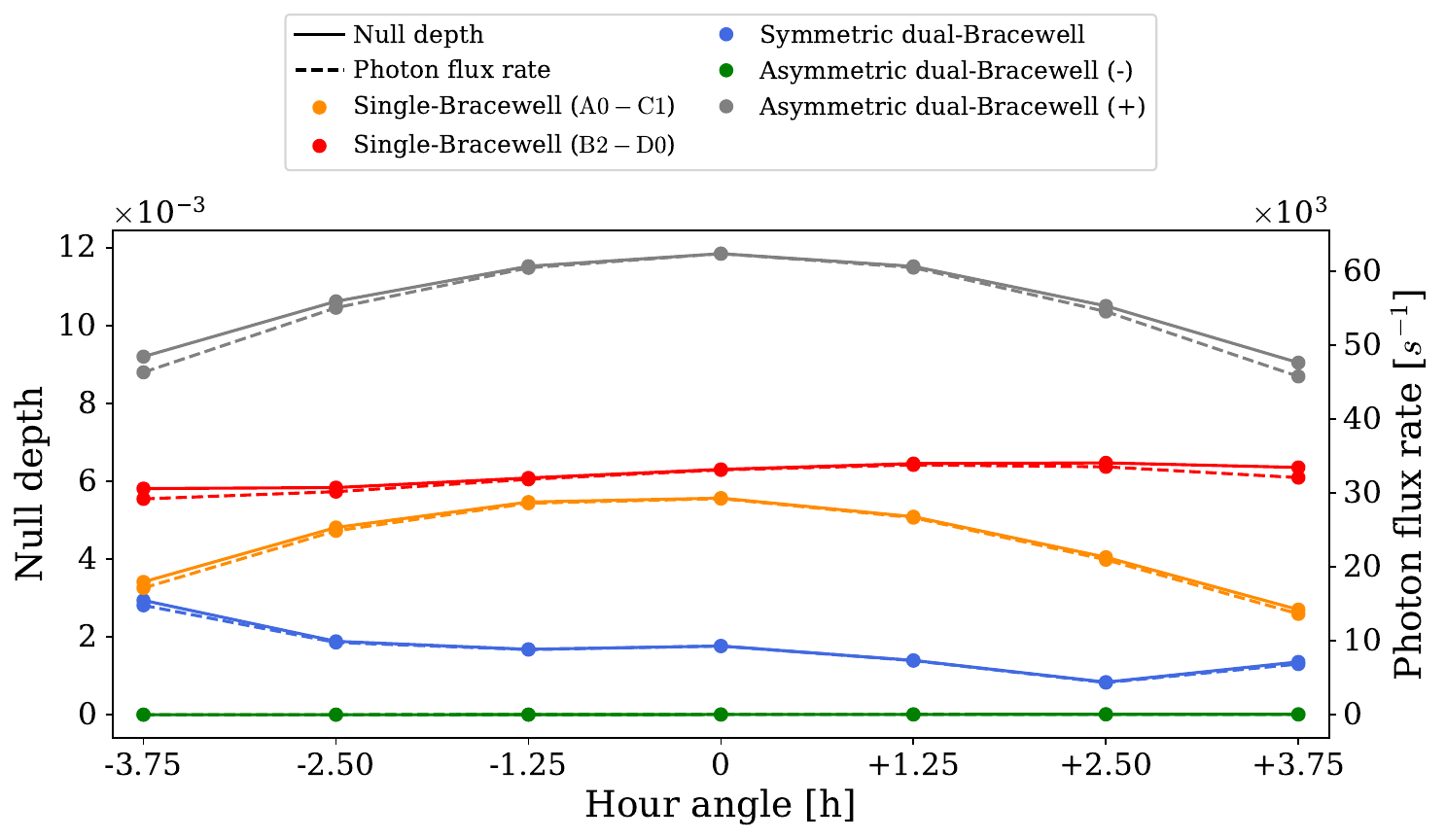}
\caption{Null depth and photon flux rate of the HEZD around \(\tau\) Ceti for the different combiner modes single-Bracewell, asymmetric dual-Bracewell (\(\pm\)), and symmetric dual-Bracewell as a function of the hour angle. Consider \hyperref[Gl9]{Eq.~\ref*{Gl9}}, \hyperref[Gl10]{Eq.~\ref*{Gl10}}, \hyperref[Gl11]{Eq.~\ref*{Gl11}}  and \hyperref[Gl12]{Eq.~\ref*{Gl12}}  for the definitions of their corresponding null depths (see \Cref{Nulling} and \Cref{combinermodes} for details).}\label{Fig4}
\end{figure}

\Cref{Fig4} illustrates the temporal evolution of the null depth and photon flux rate for the different combiner modes (single-Bracewell, asymmetric dual-Bracewell (\(\pm\)) and symmetric dual-Bracewell). The asymmetric dual-Bracewell configuration exhibits a clear performance split: the corresponding additive mode reaches non-zero values in both photon flux rate and null depth, with a maximum photon flux rate of $\approx 60 \times 10^{3}\ s^{-1}$. In contrast, the differential mode remains zero for both observables throughout the observation. This behaviour arises because this combiner mode is sensitive only to asymmetric structures, whereas our HEZD model is a symmetric ring; as a result, the subtraction of the output photon flux rates according to \hyperref[Gl8]{Eq.~\ref*{Gl8}}  yields zero signal in both the photon flux rate and the null depth. Conversely, the single-Bracewell mode exhibits larger null depths and distinct temporal behaviours. These differences arise from the distinct baselines between the telescopes (\(\mathrm{A}0-\mathrm{C}1\) and \(\mathrm{B}2-\mathrm{D}0\)), which produce different transmission maps and consequently lead to variations in both the photon flux rate and the null depth. Moreover, the previously mentioned $1\sigma$ null depth precision of $\approx 4 \times 10^{-4}$ for \(\tau\) Ceti in the symmetric dual-Bracewell mode also applies to the single-Bracewell mode and the additive asymmetric dual-Bracewell mode, with the HEZD signals lying well above this value. For the differential asymmetric dual-Bracewell mode, the $1\sigma$ null depth precision in the case of stars with similar \(L’\) band magnitude to \(\tau\) Ceti is $\approx 1 \times 10^{-4}$ (\citealt{Scott2026}).

\subsubsection{Observing wavelength}\label{wavelength}

\begin{figure}
\includegraphics[scale=0.36]{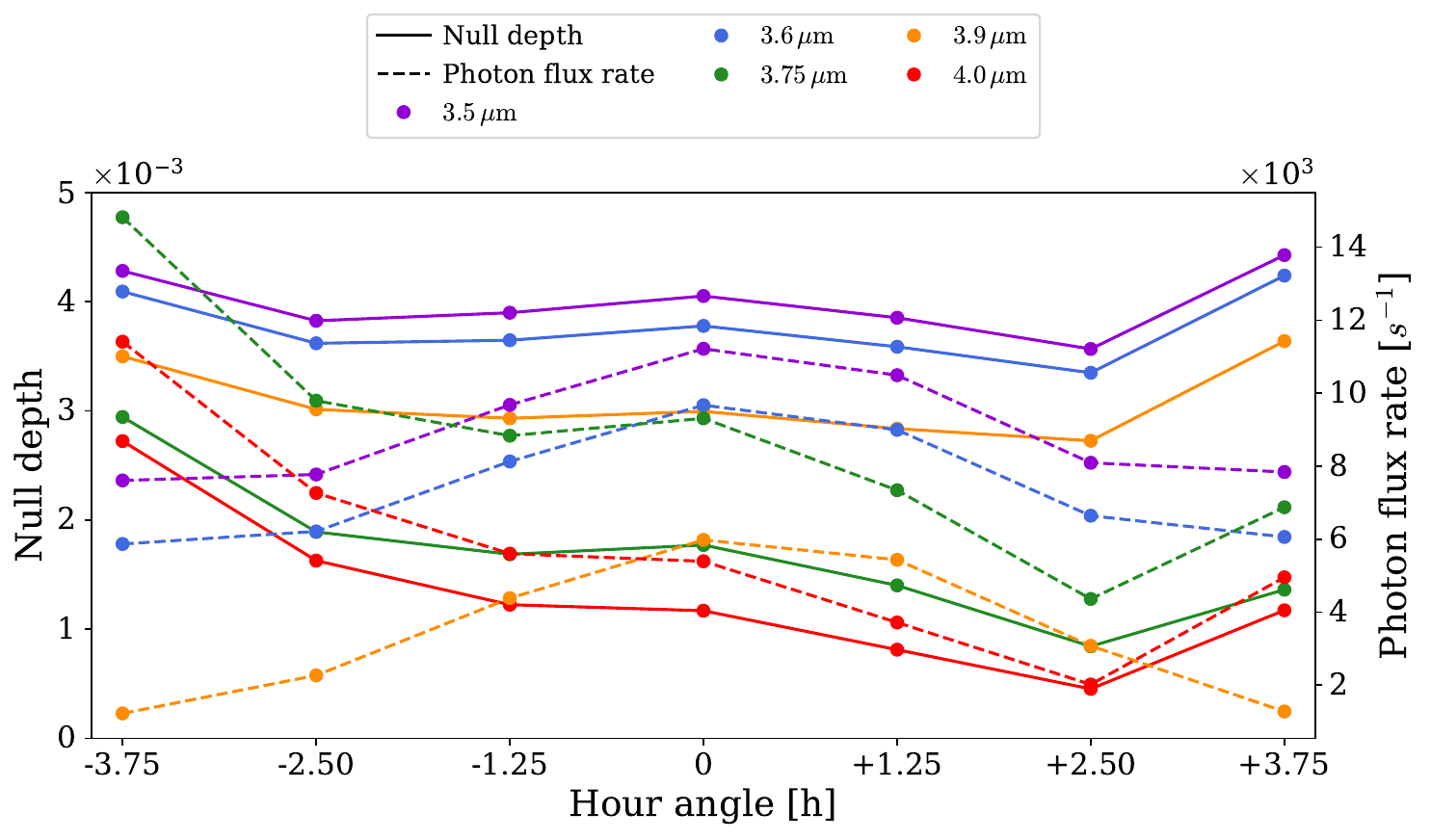}
\caption{Null depth \(N_{\mathrm{sym}}\)  and photon flux rate \(P_{\mathrm{sym}}\) of the HEZD around \(\tau\) Ceti for different observing wavelengths (see \Cref{wavelength} for details).}\label{Fig5}
\end{figure}

\Cref{Fig5} illustrates the wavelength dependence of the null depth and the photon flux rate for five wavelengths between $3.5\ \mu\mathrm{m}$ and $4.0\ \mu\mathrm{m}$. Both observables vary over the simulated observation sequence due to the wavelength-dependent transmission maps and HEZD brightness distributions. This results in spectral variations of the nulling efficiency, with some wavelengths achieving significantly larger null depths and photon flux rates than others. In particular, the shorter-wavelength end of the $L'$ band appears to be more sensitive for HEZD detection than the longer-wavelength end. At the meridian transit, the null depth spans a range from $\approx 1.1 \times 10^{-3}$ to $\approx 4.2\times 10^{-3}$, indicating that the detectability of the HEZD contribution varies across the observed wavelength range. Similarly, the photon flux rate exhibits wavelength-dependent variations, ranging from $\approx 5 \times 10^{3}$ to $\approx 11 \times 10^{3}\ s^{-1}$. Notably, the corresponding transmission maps are computed monochromatically for each central wavelength; comparing this approach to finite bandwidth maps yields a negligible $0.34\%$ root-mean-square residual over the sampled FOV, with a maximum deviation of $\approx 1\%$ due to bandwidth smearing.

\subsubsection{Telescope configuration}\label{configuration}
\begin{figure}
\centering
\includegraphics[width=\columnwidth]{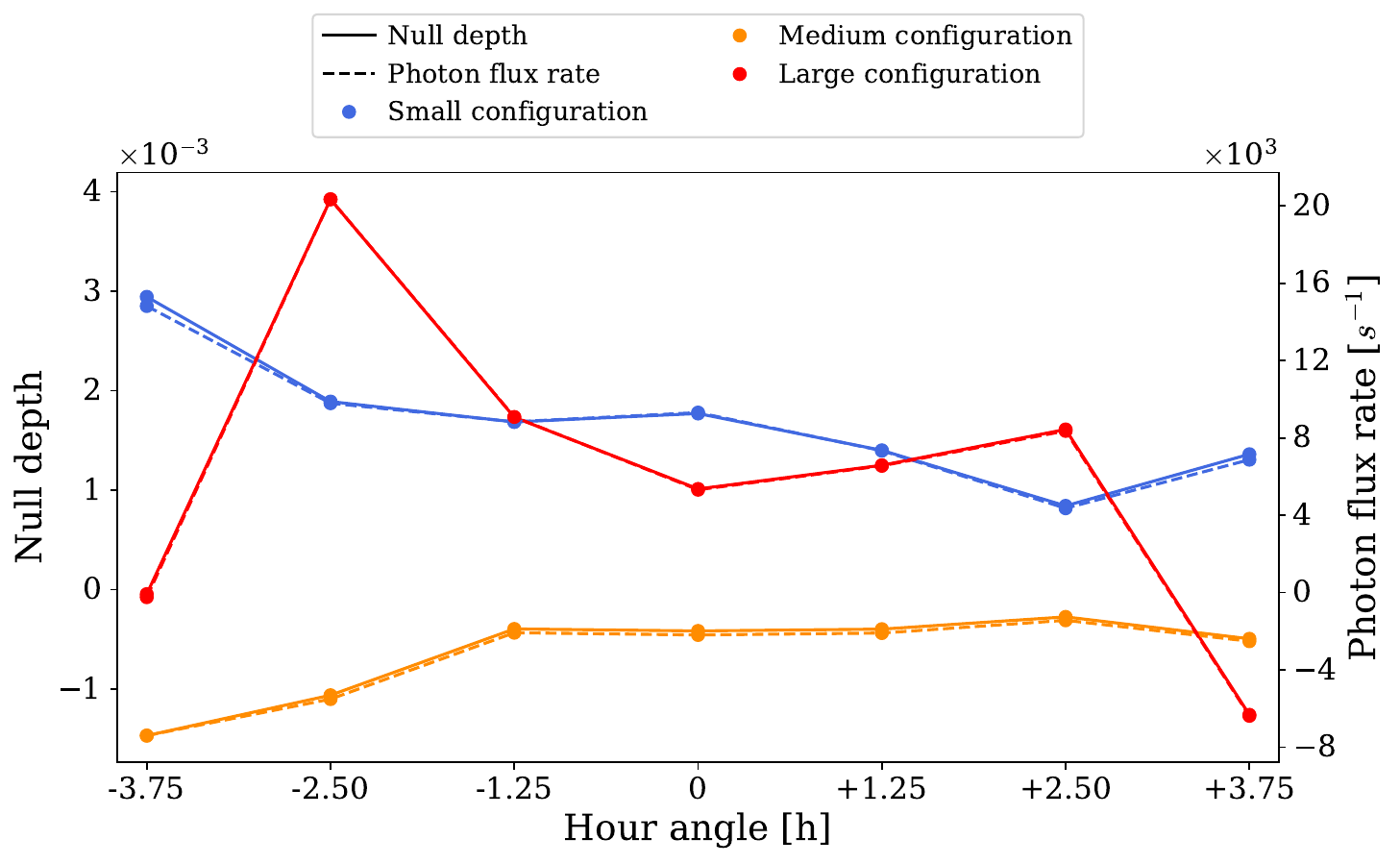}

\vspace{0.3cm}

\includegraphics[width=1.01\columnwidth]{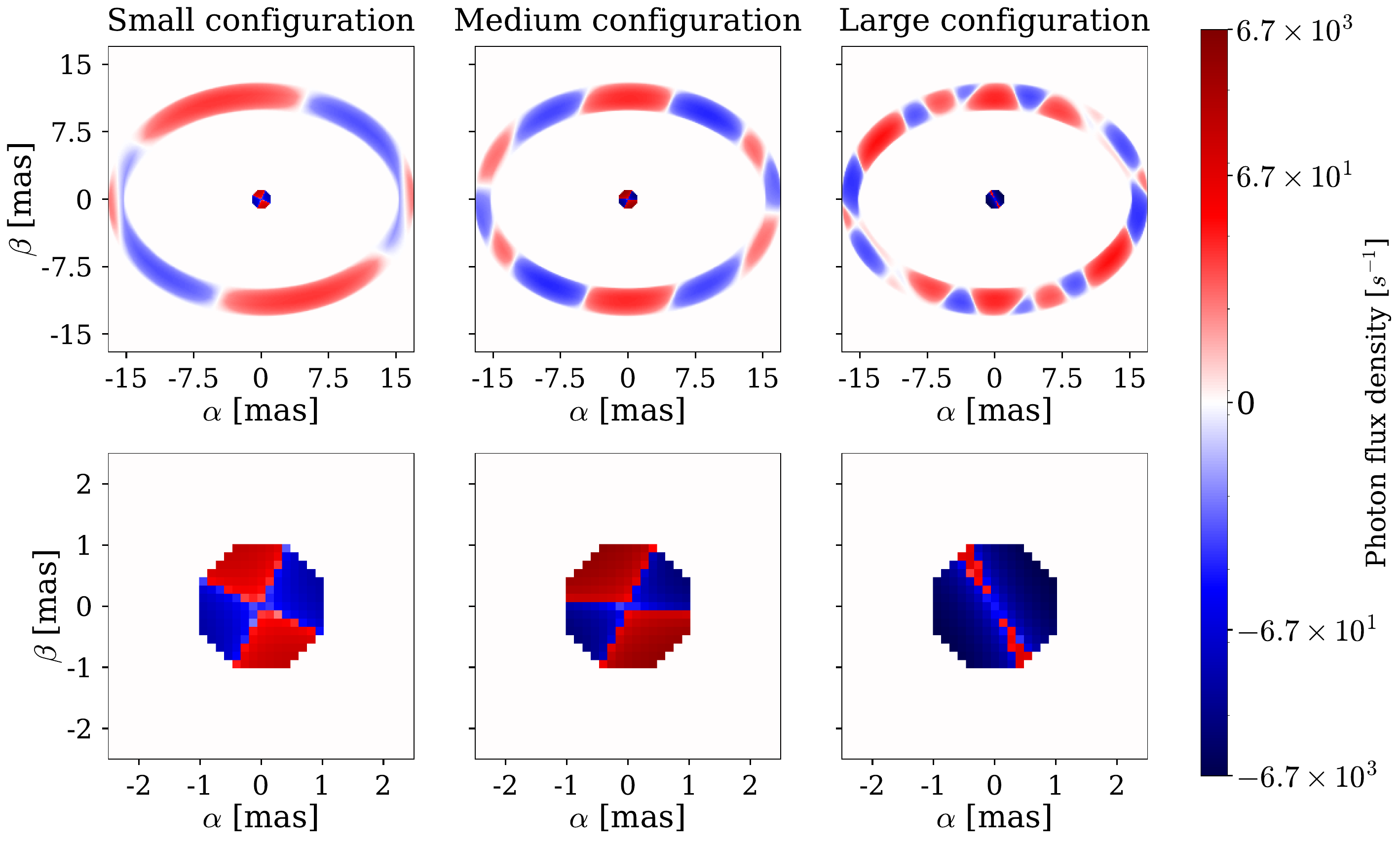}

\caption{Top: Null depth \(N_{\mathrm{sym}}\) and photon flux rate \(P_{\mathrm{sym}}\) of the HEZD around \(\tau\) Ceti for three different telescope configurations (small, medium, large) as a function of the hour angle. Mid: Corresponding photon flux density \(P\) for the meridian transit. Bottom: Close-up showing the photon flux density that arises purely from stellar leakage (see \Cref{configuration} for details).} \label{Fig6}
\end{figure}
The top plot of \Cref{Fig6} illustrates the impact of the VLTI configuration on NOTT’s observables by comparing three setups: small (baselines: $11-34$\,m), medium (baselines: $34-76$\,m), and large (baselines: $60-109$\,m), where the listed ranges indicate the minimum and maximum projected baseline lengths over the $7.5\,\mathrm{h}$ observing window. The results for the small configuration were already discussed in \Cref{referenceresults}. The predominance of negative values for both observables in the medium configuration arises from its longer baselines, which preferentially probe negative regions of the corresponding transmission maps. The null depth fluctuates between $\approx -1.5 \times 10^{-3}$ and $\approx -0.4 \times 10^{-3}$, while the photon flux rate ranges from $\approx -8 \times 10^{3}$ to $\approx -2 \times 10^{3}\ s^{-1}$. The large configuration displays a prominent peak at an hour angle of $-2.5\,\mathrm{h}$, reaching a null depth of $\approx 3.9 \times 10^{-3}$ and a photon flux rate of $\approx 20.0 \times 10^{3}\,s^{-1}$, before decreasing to negative values towards the end of the observing window.\\
\\
The photon flux density maps at the meridian transit are shown in the mid panel of \Cref{Fig6}. The small configuration, with its shorter projected baselines, produces broad transmission lobes spanning angular scales of roughly $10$ mas. By contrast, the large configuration, with significantly longer baselines, generates a high-frequency interference pattern with tightly alternating positive and negative regions. This results in strong sensitivity to baseline orientation: even small changes in the projected baselines cause rapid variations in the transmission pattern, producing the pronounced temporal modulations. 

\subsubsection{Input order}\label{input}
\begin{figure}
\centering
\includegraphics[width=\columnwidth]{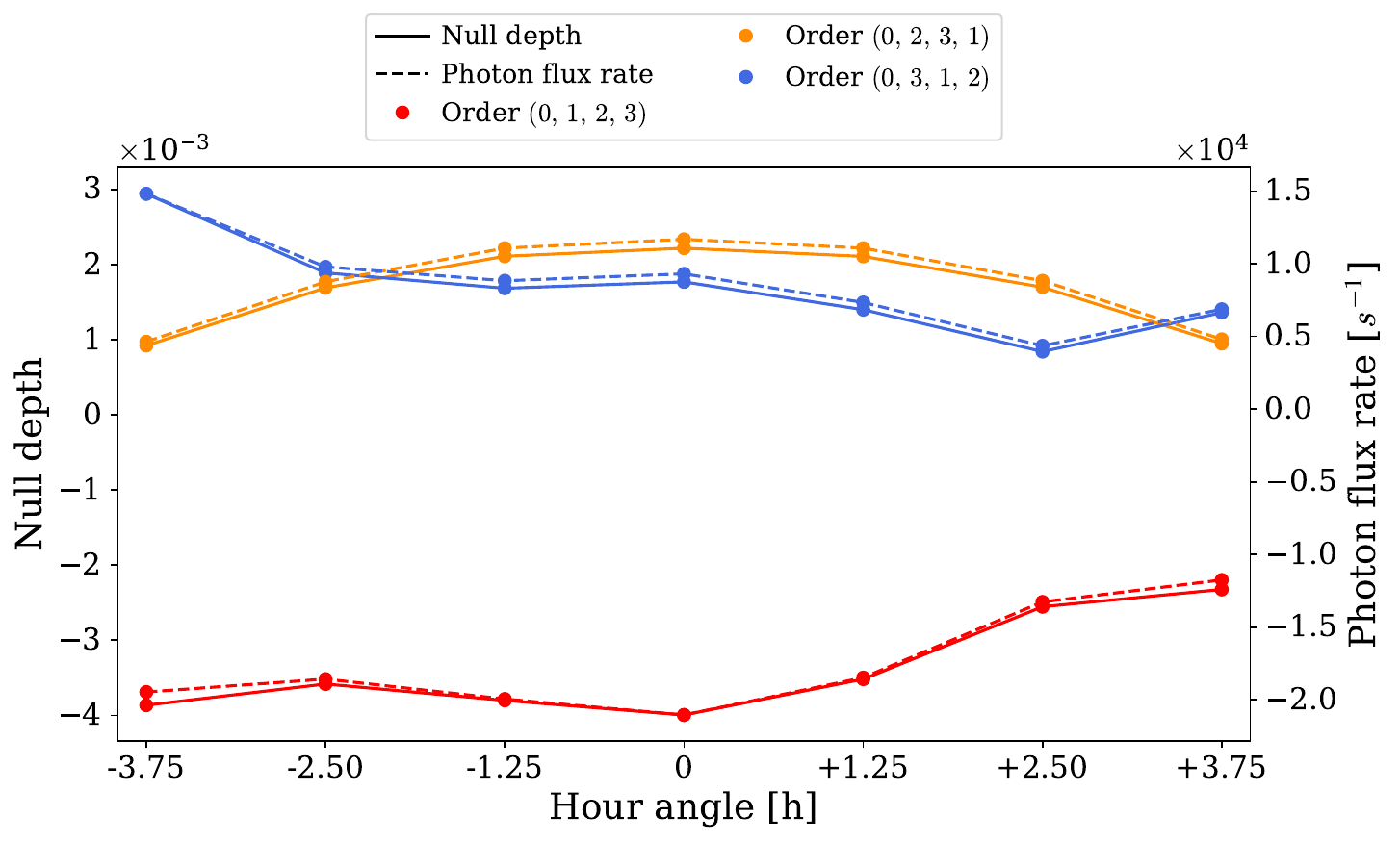}

\vspace{0.3cm}

\includegraphics[width=1.01\columnwidth]{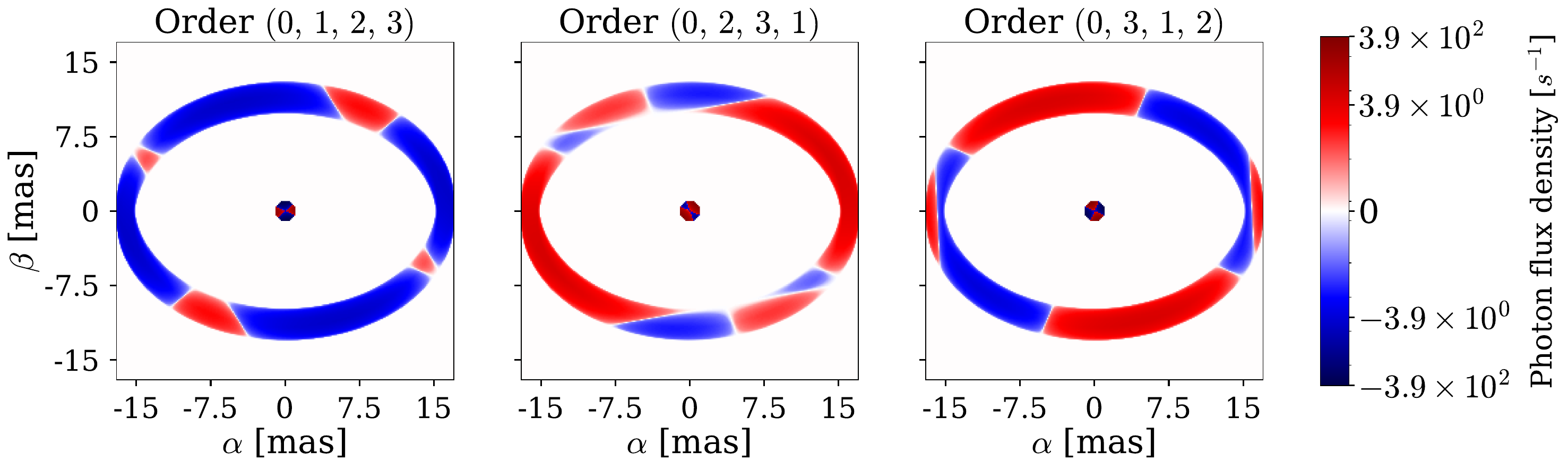}

\caption{Top: Null depth \(N_{\mathrm{sym}}\) and photon flux rate \(P_{\mathrm{sym}}\) of the HEZD around \(\tau\) Ceti for three different orders of the input telescopes ((0, 1, 2, 3), (0, 2, 3, 1), and (0, 3, 1, 2)) as a function of the hour angle. Bottom: Corresponding photon flux density \(P\) for the meridian transit (see \Cref{input} for details).} \label{Fig7}
\end{figure}

Changing the input order effectively alters the projected baselines that contribute to the null depth and photon flux rate. Thus, we illustrate in \Cref{Fig7} the impact of the different telescope input orders on NOTT’s observables by comparing three distinct ordering sequences: (0, 1, 2, 3), (0, 2, 3, 1), and (0, 3, 1, 2). The numbers correspond to the VLTI small configuration telescopes as follows: (0, 1, 2, 3) = (A0, B2, D0, C1), (0, 2, 3, 1) = (A0, D0, C1, B2), and (0, 3, 1, 2) = (A0, C1, B2, D0). The (0, 1, 2, 3) sequence produces negative null depths and photon flux rates. In contrast, the (0, 2, 3, 1) input order produces a symmetric signal peaking at the meridian transit. These differences demonstrate that the input order directly shapes the temporal behaviour of the observables, rendering some configurations significantly more sensitive to the presence of HEZD than others. Using multiple input orders during an observational campaign therefore provides additional spatial information, helping to constrain the spatial distribution of the circumstellar emission.

\subsection{HEZD parameters} \label{hezd parameters}
Starting from the parameter values of our reference model (see \Cref{table1}), we investigate the influence of selected HEZD parameters on NOTT’s observables. We first examine different radial locations of the HEZD ring (\Cref{ringradius}) and variations in the ring width (\Cref{ringwidth}). We then analyse the impact of the star–HEZD orientation, exploring how changes in ring inclination and position angle shape the expected photon flux rate and null depth (\Cref{inclinationposition}).

\subsubsection{Ring radius} \label{ringradius}

\begin{figure}
\includegraphics[scale=0.35]{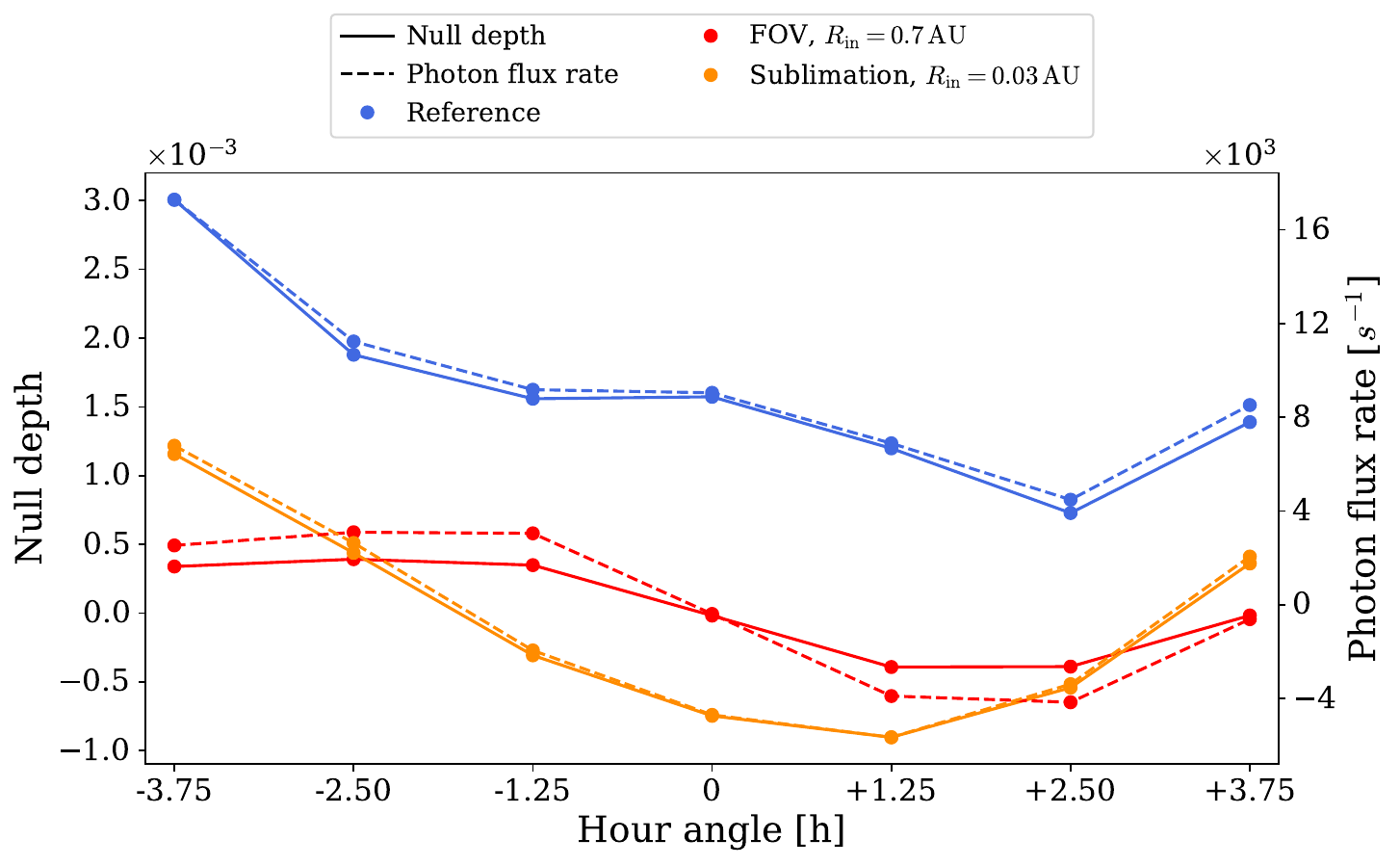}
\caption{Null depth \(N_{\mathrm{sym}}\)  and photon flux rate \(P_{\mathrm{sym}}\) of the HEZD around \(\tau\) Ceti for different inner and outer radii of the ring as a function of the hour angle (see \Cref{ringradius} for details).}\label{Fig9}
\end{figure}

The impact of the HEZD location on the simulated signal is investigated by comparing three different scenarios, as shown in \Cref{Fig9}: the sublimation case ($R_{\mathrm{in}} = 0.03\ \mathrm{au}$), corresponding to the distance between HEZD and \(\tau\) Ceti where \(0.1\ \mu\)m-sized graphite grains reach 2000 K; the reference case ($R_{\mathrm{in}} = 0.054\ \mathrm{au}$, see \Cref{referenceresults}); and the FOV case where the HEZD is located at the outer edge of NOTT's FOV ($R_{\mathrm{in}} = 0.7\ \mathrm{au}$). The results demonstrate a strong sensitivity of the observables to the proximity of the HEZD to the star. The highest photon flux rate and null depth are not obtained when the HEZD is closest to the star, but rather at the HEZD reference location. This indicates that, in the reference configuration, the HEZD is located at an angular separation that aligns well with the transmission map, making it more likely to be detected in a NOTT observation compared to the other two scenarios. By contrast, the sublimation scenario shows a diminished signal. This behaviour arises from the proximity of the HEZD to the broader dark fringes of the transmission pattern near the centre of the FOV. In the FOV case, the signal is further reduced due to the lower thermal emission at larger distances from the star and the significantly lower transmission near the edge of the FOV. This illustrates that careful selection of the interferometric configuration is essential, as an optimal regime exists in the interplay between the source extent and the setup.

\subsubsection{Ring width} \label{ringwidth}
\begin{figure}
\centering
\includegraphics[width=\columnwidth]{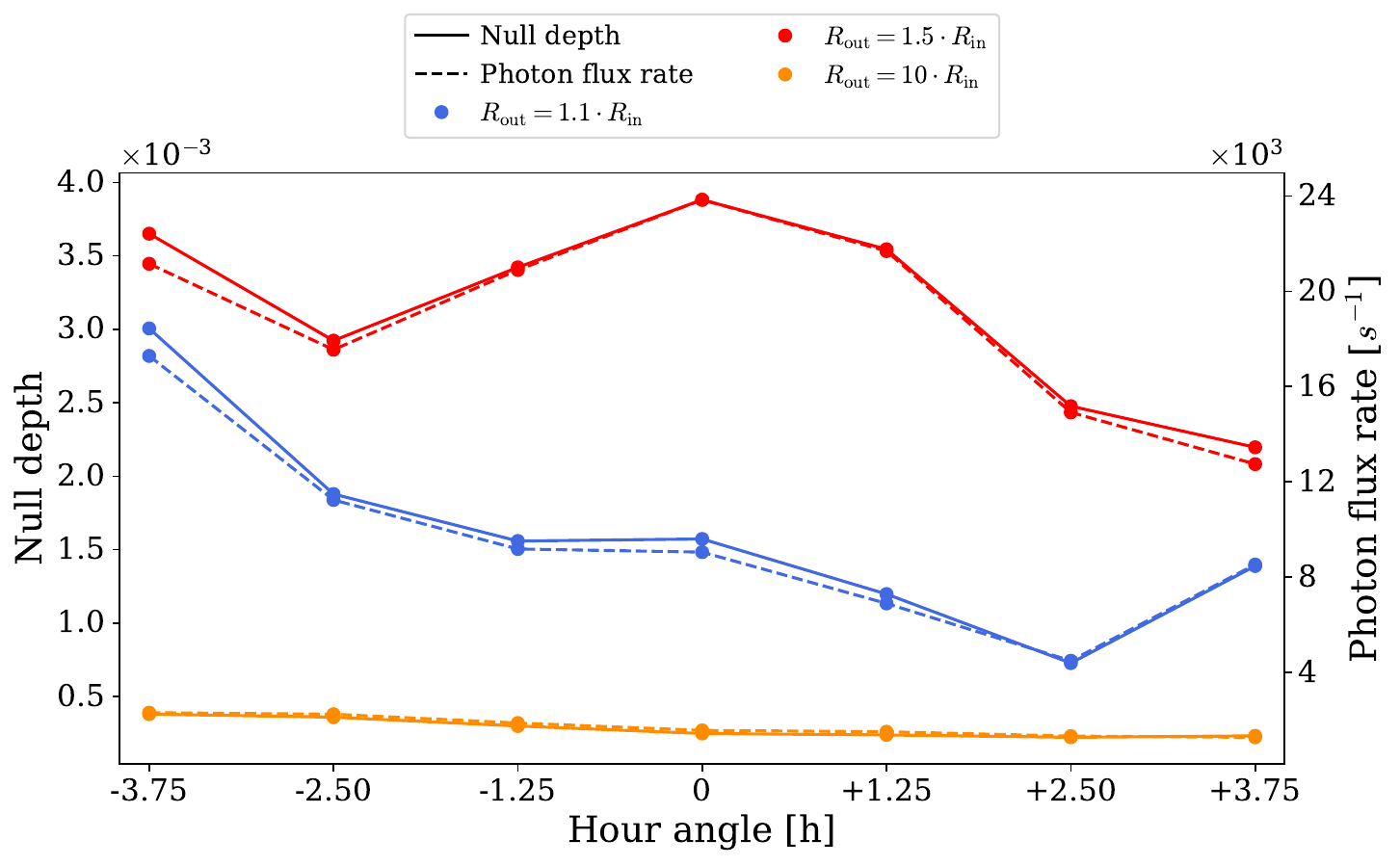}

\vspace{0.3cm}

\includegraphics[width=1.01\columnwidth]{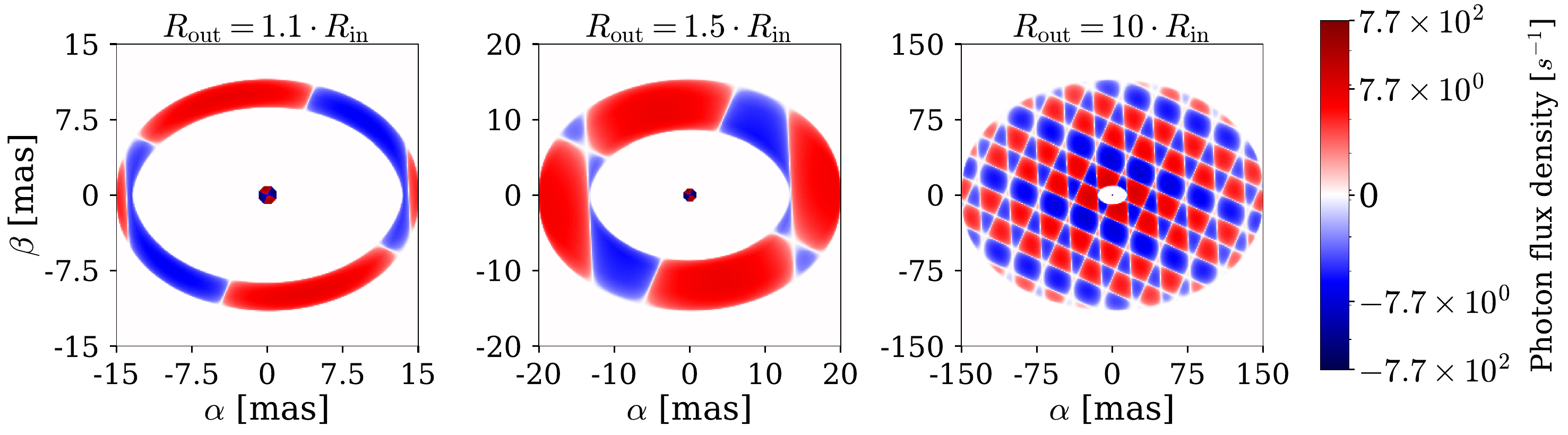}

\caption{Top: Null depth \(N_{\mathrm{sym}}\) and photon flux rate \(P_{\mathrm{sym}}\) of the HEZD around \(\tau\) Ceti for different ring widths as a function of the hour angle. Bottom: Corresponding photon flux density \(P\) for the meridian transit (see \Cref{ringwidth} for details).} \label{Fig10}
\end{figure}

\Cref{Fig10} depicts how the radial extent of the HEZD affects the observables by varying the ratio between the outer ring radius $R_{\mathrm{out}}$ and the inner ring radius $R_{\mathrm{in}}$. The top plot of \Cref{Fig10} shows that the configuration with $R_{\mathrm{out}} = 1.5 \cdot R_{\mathrm{in}}$  produces the largest photon flux rate and reaches a null depth of $\approx 3.6 \times 10^{-3}$ at meridian transit of the simulated observation. This suggests that such an intermediate ring width represents an optimal balance: the HEZD is bright enough and distributed such that it couples efficiently to the interferometric transmission pattern. This finding is consistent with the observational constraints on HEZD to date. In contrast, both the narrower ($R_{\mathrm{out}} = 1.1 \cdot R_{\mathrm{in}}$) and the extended scenarios ($R_{\mathrm{out}} = 10 \cdot R_{\mathrm{in}}$) yield lower signals.\\
\\
The bottom panel maps confirm this interplay: for the $R_{\mathrm{out}} = 1.5 \cdot R_{\mathrm{in}}$ scenario, the emission avoids destructive cancellation within the transmission lobes. In contrast, for the extended scenario ($R_{\mathrm{out}} = 10 \cdot R_{\mathrm{in}}$), the surface brightness drops significantly despite the larger angular area, leading to the observed decrease in the integrated observables.

\subsubsection{Inclination and position angle} \label{inclinationposition}

\begin{figure}
\centering
\includegraphics[width=\columnwidth]{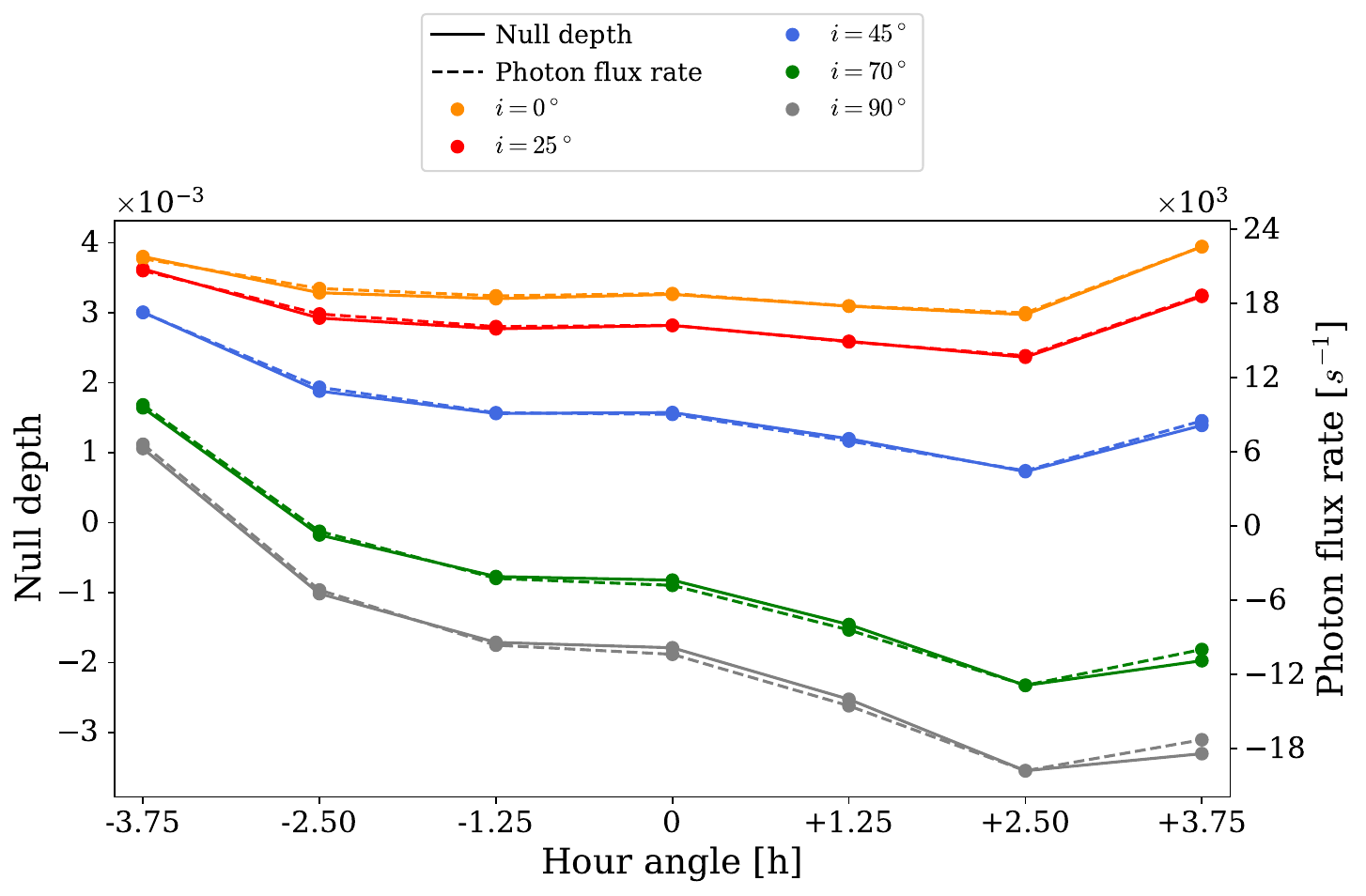}

\vspace{0.3cm} 

\includegraphics[width=1.01\columnwidth]{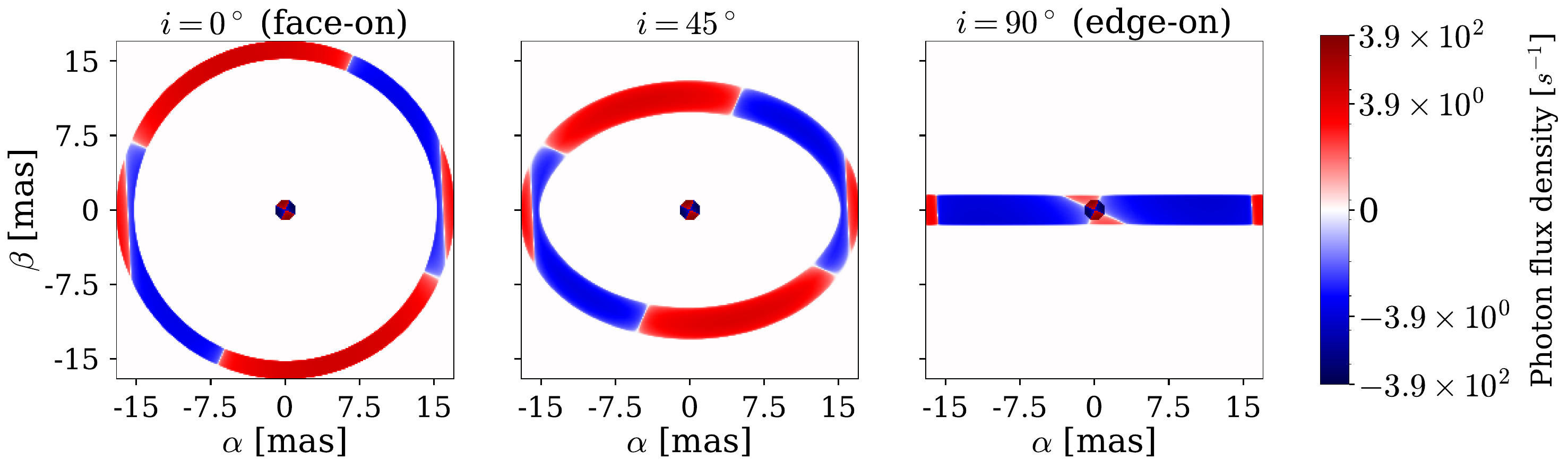}

\caption{Top: Null depth \(N_{\mathrm{sym}}\) and photon flux rate \(P_{\mathrm{sym}}\) of the HEZD around \(\tau\) Ceti for different ring inclinations as a function of the hour angle. Bottom: Corresponding photon flux density \(P\) for three selected inclinations (\(0^\circ, 45^\circ, 90^\circ\)) for the meridian transit (see \Cref{inclinationposition} for details).} \label{Fig11}
\end{figure}
\begin{figure}
\includegraphics[scale=0.29]{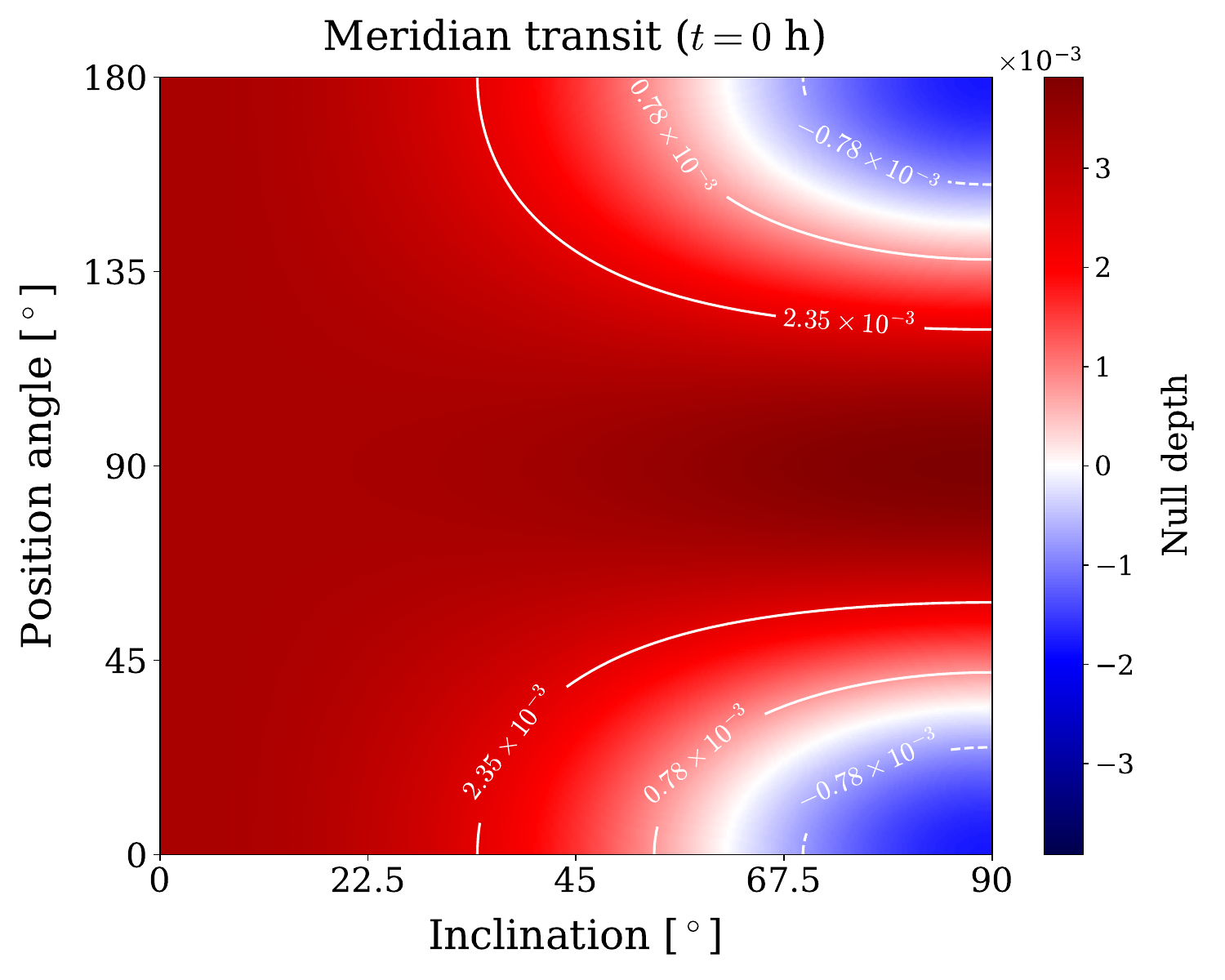}
\caption{Null depth \(N_{\mathrm{sym}}\) of the HEZD around \(\tau\) Ceti at the meridian transit for different ring inclinations and position angles. Note that inclination and position angle are partially degenerate, so different combinations of the two parameters can produce similar null-depth signatures  (see \Cref{inclinationposition} for details).}\label{Fig12}
\end{figure}

The top plot of \Cref{Fig11} illustrates the influence of the ring inclination $i$ on the observed null depth and photon flux rate, for five scenarios ranging from a face-on configuration ($i = 0^{\circ}$) to an edge-on configuration ($i = 90^{\circ}$). For a face-on ring, the circular symmetry produces only a weak variation of the observables with respect to the orientation of the projected baselines. With increasing inclination, a pronounced modulation emerges and grows in amplitude. Accordingly, the edge-on configuration exhibits the strongest variations, with the photon flux rate oscillating between $\approx 7 \times 10^{3}\,s^{-1}$ and $\approx -19 \times 10^{3}\,s^{-1}$. Increasing inclination breaks the spatial symmetry of the emission, causing the HEZD emission to be sampled by the positive and negative lobes of the instrumental transmission map. Consequently, the inclination strongly modulates the fraction of HEZD emission transmitted through NOTT: while face-on rings remain largely insensitive to baseline orientation, nearly edge-on rings experience strong suppression when the major axis aligns with the central dark fringe.\\
\\
The photon flux density maps for selected inclinations ($0^{\circ}$, $45^{\circ}$, and $90^{\circ}$) are shown in the bottom panel of \Cref{Fig11}. In the face-on case, the HEZD emission forms a circular ring that uniformly overlaps the central transmission lobes, explaining the near temporal invariance of the observables. With increasing inclination, the ring projects into an ellipse and appears as a narrow structure in the edge-on case, causing the emission to alternate between the positive and negative regions of the transmission pattern. Overall, these results demonstrate that inclination is a key parameter shaping the temporal signature of the expected HEZD signal for detections with NOTT.\\
\\
\Cref{Fig12} depicts the null depth at the meridian transit as a function of ring inclination $i$ and position angle PA of the projected ellipse. For an inclination of \(0^{\circ}\), the ring appears circular and the position angle is not defined. As the inclination increases, the influence of the position angle becomes increasingly pronounced. For strongly inclined rings (\(i \rightarrow 90^{\circ}\)), the null depth is strongly sensitive to the position angle, varying between \(\approx 3 \times 10^{-3}\) and \(\approx -3 \times 10^{-3}\). This modulation is governed by the relative alignment between the ring’s major axis and the fringe pattern of the transmission map: alignment with the central dark fringe leads to strong suppression, whereas a perpendicular orientation maximises the transmitted signal. This reflects the geometric nature of the signal: different combinations of inclination and position angle can produce similar projected brightness distributions and therefore similar null depths.

\subsection{Stellar parameters}\label{stars}

\begin{table}
\caption{Stellar parameters and coordinates of a representative F star (\(\kappa\) Tuc A), A star (Fomalhaut), and G star (\(\tau\) Ceti), whereby the astrometric data were obtained from Gaia DR3 via the ESA Gaia Archive (\citealt{Gaia}) for the observation date (2026 October 02).}
\label{table2}
\centering
\begin{tabular}{c c c c} 
 \hline
 
 Parameter & \(\kappa\) Tuc A & Fomalhaut & \(\tau\) Ceti \\ 
 \hline
 Effective temperature \(T_\star\) [K] & 6451  & 8590 & 5344      \\
 Radius \(R_{\star}\)  [mas]  & 0.37  & 1.11 & 0.98 \\
 Right ascension [\(^\circ\)] & 18.95  & 344.41 & 26.01 \\ 
 Declination [\(^\circ\)] & -68.88  & -29.62 & -15.93 \\
 Meridian transit time [LST]  & 01:16  & 22:58 & 01:44 \\
 Distance to observer \(d_{\mathrm{obs}}\) [pc] & 20.99  & 7.70 & 3.65\\
 \hline
\end{tabular}
\end{table}

\begin{figure}
\includegraphics[scale=0.36]{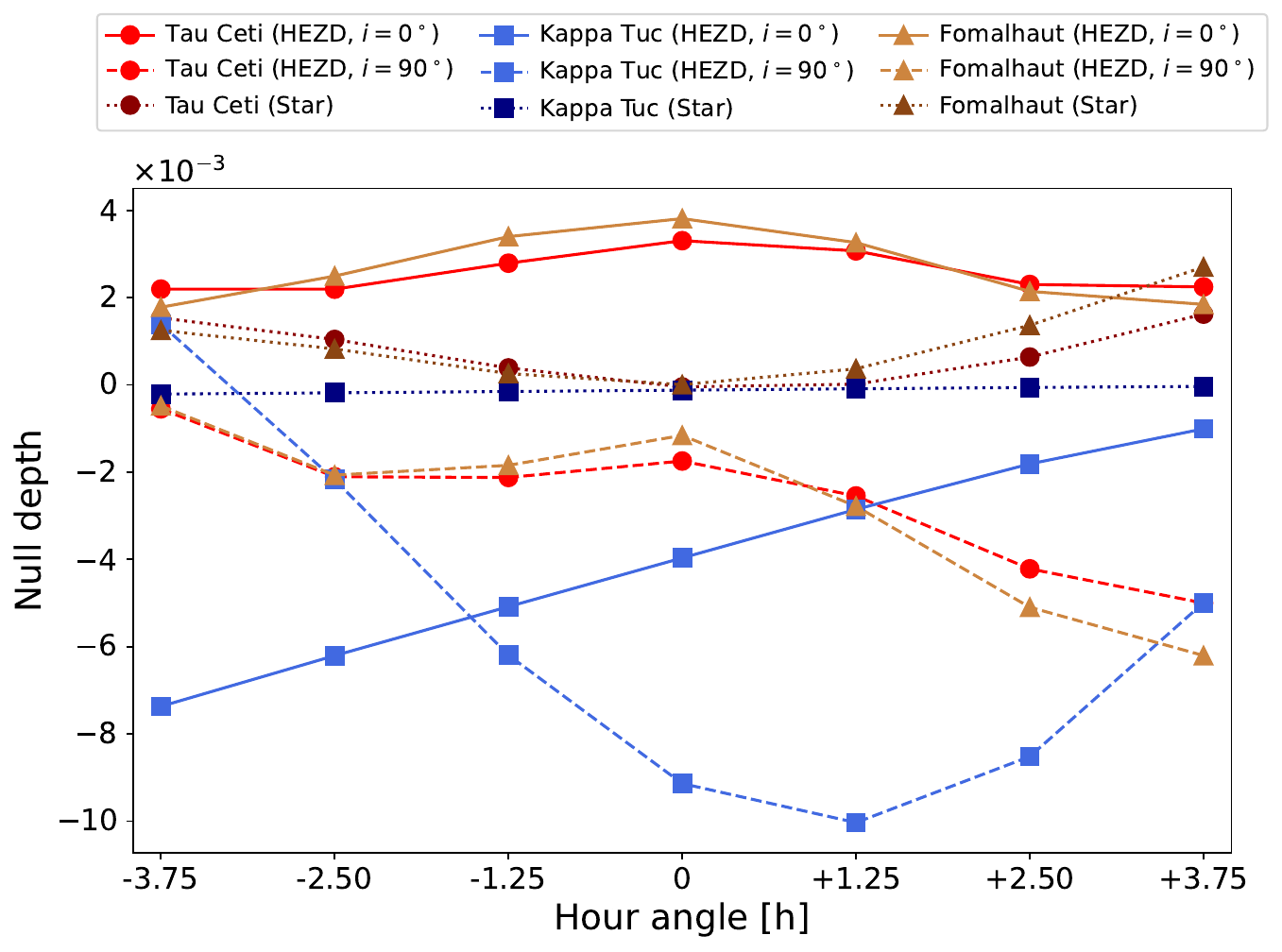}
\caption{Null depth \(N_{\mathrm{sym}}\) for \(\kappa\) Tuc A, Fomalhaut, and \(\tau\) Ceti with HEZD as a function of the hour angle. The HEZD mass is adjusted for each star to maintain a constant dust-to-star flux ratio of $1\,\%$ (see \Cref{stars} for details).}\label{Fig13}
\end{figure}

While our stellar model has so far been adapted exclusively to mimic \(\tau\) Ceti, we now extend the study to two additional representative stellar systems, Fomalhaut and \(\kappa\) Tuc A, to investigate their potential HEZD signatures in the context of prospective NOTT observations (see \Cref{table2}). The A3V star Fomalhaut (HD 216956), at a distance of $7.70$\,pc (\citealt{Reyle}), with an effective temperature of $8590$\,K (\citealt{Mamajek}) and with a radius of \(1.11\) mas (\citealt{Absil2009}), is a prime target of HEZD research in the NIR and MIR \citep{Absil2009, Lebreton, Mennesson, Ollmann2025}. Moreover, \(\kappa\) Tuc A (HD 7788 A), classified as an F6IV star \citep{Tokovinin2020} at a distance of $20.99$\,pc \citep{Gaia}, with an effective temperature of $6451$\,K \citep{Fuhrmann2017} and with a radius of \(0.37\) mas (\citealt{Kirchschlager2020}), represents another key target for HEZD observations. Observations of \(\kappa\) Tuc A were obtained with MATISSE (Multi AperTure mid-Infrared SpectroScopic Experiment; \citealt{Lopez, Lopez2022}), a second-generation VLTI instrument operating in the $L$, $M$, and $N$ bands, and revealed a significant \(L\) band excess, attributed to the presence of HEZD \citep{Kirchschlager2020}. HEZD emission was also inferred with PIONIER in the \(H\) band in 2012 and 2014, but not in 2013, indicating temporal variability of the signal \citep{Ertel, Ertel2016}. Recently, observations with MATISSE and the instrument GRAVITY (\citealt{Gravity}) revealed a stellar companion close to \(\kappa\) Tuc A that could have caused the 2012 and 2014 HEZD signals \citep{Stuber2026}. NOTT is suited to distinguish between these scenarios, for instance, by using the additive and differential dual-Bracewell modes to search for symmetric (dust) and asymmetric (companion star) structures.\\
\\
For \(\tau\) Ceti, we retain the reference HEZD model used throughout this study. For \(\kappa\) Tuc~A and Fomalhaut, we adapt the stellar distance and astrometric coordinates to the respective systems and adopt the inner and outer HEZD radii from \citet{Kirchschlager2020} and \citet{Ollmann2025}, respectively. The HEZD mass is then adjusted for each system to maintain the fixed dust-to-star flux ratio of \(1\,\%\) at \(3.75\,\mu\mathrm{m}\). In this context, we follow \citet{Kirchschlager2020} and adopt amorphous carbon as the dust species for \(\kappa\) Tuc~A, while graphite is assumed for Fomalhaut and \(\tau\) Ceti. \Cref{Fig13} illustrates NOTT's expected signal for the three representative stars, showing the temporal evolution of the null depth. Comparing the signal produced by the star alone with the signal produced by the HEZD alone for different inclinations clearly illustrates the impact of the HEZD. While the stellar leakages of the individual stars are very similar, the HEZD signal generates a distinct temporal signature for each system that can be used for observational identification. For all three stars, the edge-on configurations induce strong temporal modulation and predominantly negative null depths. This is most prominent for \(\kappa\) Tuc~A, where the null depth decreases from \(\approx 1 \times 10^{-3}\) at the beginning of the observation to \(\approx -10 \times 10^{-3}\) shortly after the meridian transit. This behaviour demonstrates that the shape of the HEZD signal for the selected stars depends on the celestial coordinates and distance to the observer, as well as on the HEZD ring radius.

\subsection{Expected NOTT results for \(\kappa\) Tuc A} \label{Kappatuc}

\begin{table}
\caption{Evaluated parameter space and the corresponding number of linearly distributed values for HEZD around \(\kappa\) Tuc A, modelled as a ring. The parameter ranges for the inner and outer ring radii were derived from the $1\sigma$ confidence intervals of the MATISSE data analysed in \citet{Kirchschlager2020}, covering 22 observing wavelengths in the range of \(3.37-3.85\,\mu\)m.}
\label{table3}
\centering
\begin{tabular}{c c c} 
 \hline
 Parameter & Explored range & Number of values \\ 
 \hline
 Inner ring radius \(R_{\mathrm{in}}\) [au] & \(0.029-1.07\) & 10000 \\ 
 Outer ring radius \(R_{\mathrm{out}}\) [au] & \(0.032-1.18\)& 10000 \\
 Ring inclination \(i\) [\(^\circ\)] & \(0-90\) & 1000 \\
 Position angle PA  [\(^\circ\)] & \(0-180\)  & 1000    \\
 \hline
\end{tabular}
\end{table}

\begin{figure}
\includegraphics[scale=0.34]{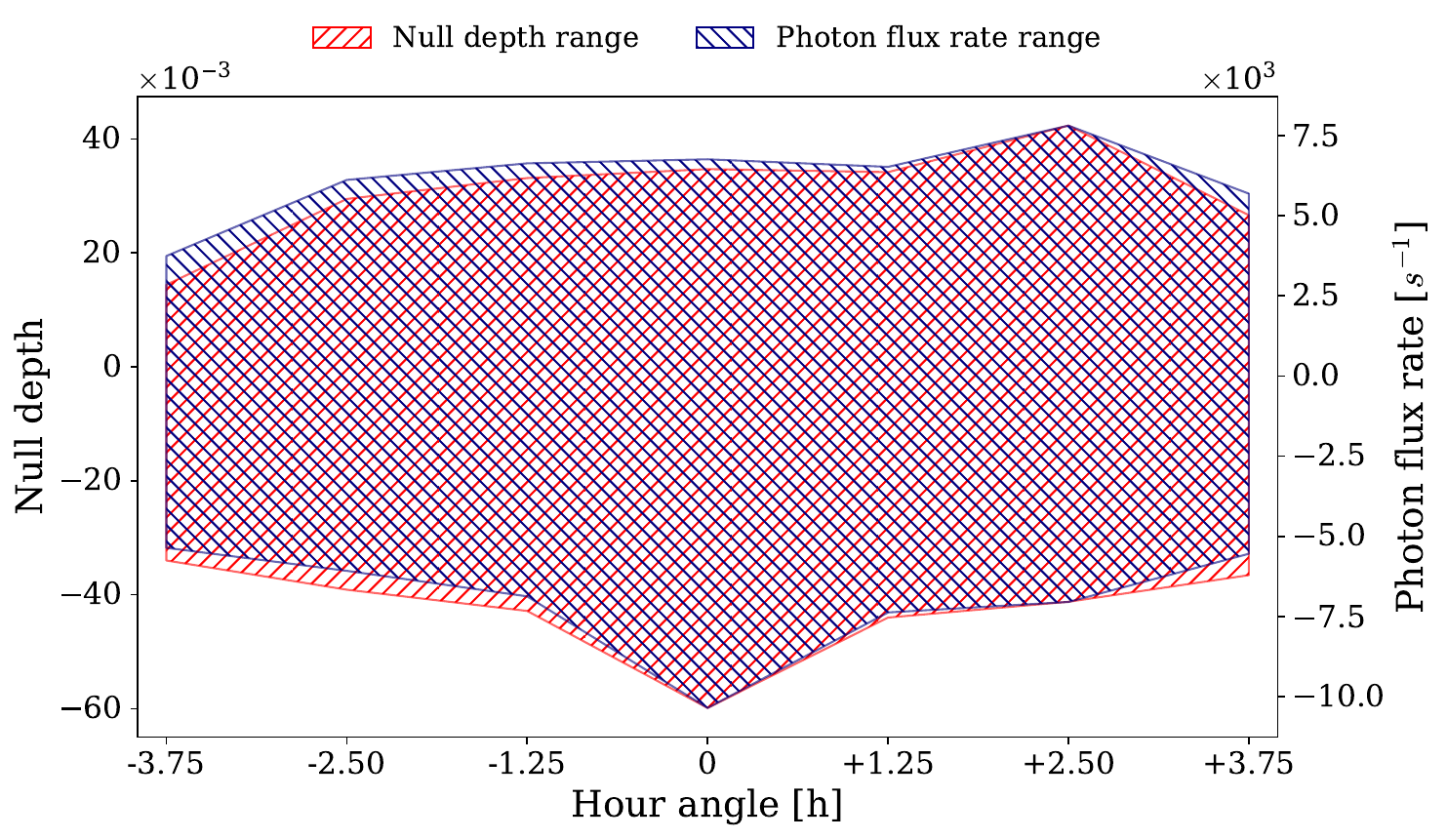}
\caption{Range of the possible null depth \(N_{\mathrm{sym}}\) and photon flux rate \(P_{\mathrm{sym}}\) of the HEZD around \(\kappa\) Tuc A as a function of the hour angle for inner and outer ring radii consistent with the \(1\sigma\) parameter intervals of \(R_{\mathrm{in}}\) and \(R_{\mathrm{out}}\) determined by \citealt{Kirchschlager2020} (see \Cref{table3}). The values are computed for 22 observing wavelengths in the range \(3.37-3.85\,\mu\)m (see \Cref{Kappatuc} for details).}\label{Fig14}
\end{figure}

Using the constraints on the inner and outer ring radii derived from the \(L\) band MATISSE observations described above, we explore the range of null depths and photon flux rates expected from HEZD around \(\kappa\) Tuc A. Alternative interpretations of this \(L\) band excess, such as a companion (\citealt{Tsishchankava2026}), are disfavoured by \citet{Stuber2026} and thus are not considered in the following analysis.\\
\\
\Cref{Fig14} presents the full range of expected null depths and photon flux rates for \(\kappa\) Tuc~A with HEZD, exploring the \(1\sigma\) confidence intervals reported by \citet{Kirchschlager2020} for the inner and outer ring radii together with variations in inclination and position angle (see \Cref{table3}). Over this parameter space, the null depth ranges from \(\approx 42 \times 10^{-3}\) to \(\approx -60 \times 10^{-3}\), while the photon flux rate ranges from \(\approx 8 \times 10^{3}\) to \(\approx -10 \times 10^{3}\,s^{-1}\); this wide dispersion reflects the substantial radial uncertainties, spanning \(0.029-1.18\)\,au. However, the parameter space shown in \Cref{Fig14} is not uniformly representative of the observational constraints: since the radial constraints stem from a fit to the MATISSE data, the most probable observables correspond to the best-fitting inner and outer radii, whose signatures are shown explicitly in \Cref{Fig13} and span a markedly narrower range of \(\approx -10 \times 10^{-3}\) to \(\approx 1 \times 10^{-3}\). This contrast demonstrates that the HEZD model for \(\kappa\) Tuc~A can in principle produce a broad range of observable signatures relevant for future NOTT measurements.

\section{Discussion} \label{discussion}
In \Cref{limits}, we evaluate the detectability of HEZD with NOTT and discuss the limitations imposed by parameter degeneracies. Subsequently, in \Cref{sec:generality_discussion}, we outline how the findings derived for \(\tau\) Ceti can be generalised to other target stars and different HEZD configurations.

\subsection{Detection limits and parameter degeneracies}\label{limits}
The parameter investigation presented in this study demonstrates the potential of NOTT for detecting and characterising HEZD. 
\\
As noted in \Cref{referenceresults}, the \(1\sigma\) null depth precision for stars with an \(L'\) band magnitude similar to that of \(\tau\) Ceti in NOTT’s symmetric dual-Bracewell mode is \(\approx 4 \times 10^{-4}\). Throughout \Cref{results}, the HEZD signal is shown to be well above this threshold in most cases. A key strength of NOTT lies in its differential measurement concept, which combines destructive outputs to form a stable Kernel null. This approach provides intrinsic robustness against instrumental and atmospheric perturbations, such as wavefront, tip-tilt, and piston errors. Consequently, the precision is largely set by photon noise from both the target and the background, meaning that the achievable performance depends primarily on the target's magnitude and the background brightness rather than on its spectral type \citep{Scott2026}.
Existing nulling instruments operating in the $N$ band, such as KIN and LBTI, are typically limited to null depths of a few $10^{-4}$--$10^{-3}$, primarily due to residual background systematics \citep{Colavita2009}. Note that the null depths of NOTT's dual-Bracewell modes are defined differently from those used in earlier studies and are therefore not directly comparable.
While geometric stellar leakage constitutes a source of uncertainty dependent on the stellar angular diameter and target distance \citep{Scott2026}, independent measurements of the stellar diameter allow for a partial calibration. This reduces parameter degeneracies and enables robust data analysis strategies, ranging from modelling the isolated HEZD emission after stellar subtraction to a joint fit of the star-HEZD system using informed priors.
The temporal modulation of the observables is driven by the interplay between the instrumental setup and the HEZD geometry. As Earth’s rotation causes the fringes of the transmission map to sweep across the target, the telescope configuration and the permutation of the input beams dictate the instrumental response. Simultaneously, the inclination and position angle of the HEZD emerge as the dominant geometric drivers. Increasing inclination breaks the spatial symmetry of the emission, causing the thermal flux to alternately couple into the positive and negative lobes of the instrumental transmission map (\Cref{Fig11,Fig12}).\\
However, as illustrated in \Cref{Fig12}, distinct combinations of inclination and position angle can produce nearly indistinguishable temporal signatures, preventing a unique reconstruction of the HEZD geometry from NOTT data acquired in a single observing mode. Resolving these parameter degeneracies requires dedicated follow-up studies based on synthetic NOTT data sets that include realistic noise and parameter retrieval, for instance using Bayesian sampling techniques. In particular, the impact of the beam-combiner mode (symmetric or asymmetric dual-Bracewell, or single-Bracewell) on the trade-off between robustness and spatial information content will be investigated in such a follow-up study.

\subsection{Generalisation of the results beyond \(\tau\) Ceti}
\label{sec:generality_discussion}
The results presented in \Cref{results} are illustrated mainly for \(\tau\)~Ceti and a specific set of baselines. While the detailed shape of the null depth and photon flux rate curves as a function of hour angle is target-specific, the underlying geometric relations can be generalised.\\
\\
The characteristic level of the null depth for a given dust-to-star flux ratio is expected to be similar across different targets, provided the angular scale of the HEZD emission (ring radius and ring width) matches the spatial scale probed by the interferometer. This spatial relationship is the determining factor, rather than the physical size of the HEZD emission alone. This is evident in the comparison of \(\tau\)~Ceti, \(\kappa\) Tuc A, and Fomalhaut under an equivalent HEZD configuration (\Cref{stars}), where differences in the resulting null depth signatures stem from their distinct angular scales and how these couple to the selected baselines.
Consequently, rather than adopting a fixed baseline configuration for all targets, baselines should be tailored to the expected angular extent of the circumstellar emission. Changes in the angular scale can, to a large extent, be accommodated by selecting appropriately matched baselines.\\
The absolute brightness of the target introduces a separate consideration: the null depth is primarily independent of absolute stellar brightness for a fixed dust-to-star flux ratio, whereas the photon flux rate scales directly with stellar flux. For sufficiently faint targets, observations become limited by photon statistics or instrumental background, making the achievable sensitivity dependent on the magnitude-dependent performance of the instrument.
Taken together, these considerations show that while the exact time-domain signatures are specific to the modelled \(\tau\)~Ceti configuration, the fundamental dependencies on dust-to-star flux ratio, angular scale, and stellar brightness can be robustly applied to other targets. The simulations therefore provide a framework for interpreting how the expected HEZD signal changes with target properties and for selecting suitable interferometric configurations in future surveys.

\section{Summary and outlook} \label{Summary}
In this study, we have assessed the scientific potential of the nulling interferometer NOTT for the detection and characterisation of HEZD. Using the end-to-end nulling simulator SCIFYsim in combination with the debris disc simulator DMS, we modelled the HEZD as an optically thin dust ring based on a reference model for the star \(\tau\) Ceti. We systematically explored the impact of instrumental parameters (including telescope configuration, combiner modes, and input permutations) and HEZD properties (such as ring inclination, position angle, inner and outer radii, and ring width) on the measurable photon flux rate and null depth, assuming that the stellar contribution can be independently constrained and calibrated through dedicated calibrator observations and reliable estimates of the stellar diameter. The main conclusions are as follows:
\begin{enumerate}
\item For the range of HEZD and instrumental parameters considered, many of the HEZD null depths remain above NOTT's expected \(1\sigma\) null depth precision for stars with an \(L'\) band magnitude comparable to that of \(\tau\) Ceti. However, intrinsic parameter degeneracies, for example, between ring inclination and position angle, limit the ability to uniquely constrain HEZD properties from NOTT observations alone (see \Cref{referenceresults}, \Cref{inclinationposition}, and \Cref{discussion}).
\item Different telescope configurations and input permutations probe complementary information about the system, as reflected in their distinct impact on the null depth and photon flux rate. This further emphasises that careful configuration selection is required, since an optimal regime emerges from the interplay between the physical extent of the source and the instrumental setup. Moreover, exploring different observational schemes can provide additional insights (see \Cref{configuration}, \Cref{input}).
\item In the case of \(\kappa\) Tuc A, we adopt the $1\sigma$ confidence intervals of the HEZD parameters derived from a previous \mbox{MATISSE} data analysis. These parameter ranges are used as input for the NOTT simulations to estimate the corresponding ranges of expected null depths and photon flux rates. This approach allows for a consistency assessment between the HEZD properties inferred from MATISSE and potential future NOTT observations. If forthcoming NOTT measurements yield null depths outside the simulated ranges, this would indicate that NOTT provides different constraints. In that case, the HEZD model inferred from the MATISSE data may require refinement (see \Cref{Kappatuc}).
\item The detailed temporal signature of the null depth and photon flux rate is target-specific, as it depends on the target geometry and the projected interferometric baselines. However, the characteristic behaviour is governed by the angular scale of the HEZD relative to the spatial scale of the interferometric transmission pattern. Consequently, the impact of different target distances or HEZD sizes can, to a large extent, be accommodated by adapting the telescope configuration to the corresponding angular scale. This provides a general guideline for selecting suitable baseline configurations when extending NOTT observations to previously unexplored targets (see \Cref{sec:generality_discussion}).
\end{enumerate} 
The sensitivity of NOTT may provide complementary constraints on HEZD parameter values around known HEZD systems and potentially facilitate the investigation of additional candidate systems. However, the extent to which such constraints can be achieved in practice will depend on instrumental performance, atmospheric conditions, and calibration systematics, which are not fully accounted for in the present performance analysis. Repeated observations may also provide an opportunity to investigate temporal variability in HEZD emission (\citealt{Ertel2016}; \citealt{Nunez}), although the sensitivity to such variations will need to be assessed under realistic observing conditions. Finally, joint data reduction remains a challenge that upcoming data formats aim to address.

\section*{Acknowledgements}

 This work is supported by the Research Unit FOR 2285 “Debris Disks in Planetary Systems” of the Deutsche Forschungsgemeinschaft (DFG). KO and SW acknowledge the DFG for financial support under contract WO 857/15-2. JPS, TAS, and SE are supported by the National Aeronautics and Space Administration (NASA) through Grant No. 80NSSC23K1473; TAS is also supported through NASA Grant No. 80NSSC23K0288. RL and the Self-Calibrated Interferometry for Exoplanet Spectroscopy (SCIFY) project received funding from the European Research Council (ERC) under the European Union's Horizon 2020 research and innovation programme (grant agreement No. 866070). SCIFYsim is a development carried out in the context of the SCIFY project. FK has received funding from the European Research Council (ERC) under the European Union's Horizon 2020 research and innovation programme DustOrigin (ERC-2019-StG-851622) and was supported by a DFG Heisenberg Fellowship. TDP is supported by a UK Research and Innovation (UKRI) Stephen Hawking Fellowship and a Warwick Prize Fellowship, the latter made possible by a generous philanthropic donation. This research was supported in part through high-performance computing resources available at the Kiel University Computing Centre. This work has made use of data from the European Space Agency (ESA) mission Gaia (https://www.cosmos.esa.int/gaia), processed by the Gaia Data Processing and Analysis Consortium (DPAC, https://www.cosmos.esa.int/web/gaia/dpac/consortium). Funding for the DPAC has been provided by national institutions, in particular the institutions participating in the Gaia Multilateral Agreement. ChatGPT (OpenAI) was used to assist with grammar and language editing; all scientific content and interpretations are the responsibility of the authors.

\section*{Data Availability}
The simulation data underlying this article will be made available upon reasonable request.







\appendix

\section{Theoretical foundation for NOTT simulations} \label{app:math_details}

For \(N\) apertures, each subaperture \(k\) is characterised by its position \((x_k, y_k)\) in the plane perpendicular to the line of sight between target and observer, as well as by its aperture radius \(r_k\). The complex field amplitude \(V_k\) generated in aperture \(k\) by a source located at the angular position \((\alpha_i, \beta_j)\) is denoted as
\begin{equation} \label{Gl1}
V_k \propto r_k \, e^{2\pi i (x_k \alpha_i + y_k \beta_j)/\lambda},
\end{equation}
where \(i\) and \(j\) denote the pixel indices of the discretised angular coordinates \(\alpha_i\) and \(\beta_j\). The beam combination is described by a complex combiner matrix \(U\) (\citealt{Guyon2013}), which connects the input amplitudes \(V_k\) to the output vectors \(W_m\) (\(m\) labels the respective output) via
\begin{equation} \label{Gl2}
W = U V.
\end{equation}
For NOTT, this combination follows a double-Bracewell scheme (\(N=4\)) for the eight outputs. In a simplified dispersion-free version, the matrix \(U\) has the form
\begin{equation} \label{Gl3}
U\propto
\left[
\begin{matrix}
\sqrt{b} & 0 & 0 & 0 \\
0 & \sqrt{b} & 0 & 0 \\
\frac{\sqrt{2}\,\sqrt{1 - b}}{2} & \frac{\sqrt{2}\,\sqrt{1 - b}}{2} & 0 & 0 \\
\frac{\sqrt{1 - b} e^{i \phi_{0}}}{2} & -\frac{\sqrt{1 - b} e^{i \phi_{0}}}{2} & \frac{\sqrt{1 - b} e^{i \phi_{1}}}{2} & -\frac{\sqrt{1 - b} e^{i \phi_{1}}}{2} \\
\frac{\sqrt{1 - b} e^{i \phi_{0}}}{2} & -\frac{\sqrt{1 - b} e^{i \phi_{0}}}{2} & -\frac{\sqrt{1 - b} e^{i \phi_{1}}}{2} & \frac{\sqrt{1 - b} e^{i \phi_{1}}}{2} \\
0 & 0 & \frac{\sqrt{2}\,\sqrt{1 - b}}{2} & \frac{\sqrt{2}\,\sqrt{1 - b}}{2} \\
0 & 0 & \sqrt{b} & 0 \\
0 & 0 & 0 & \sqrt{b}
\end{matrix}
\right],
\end{equation}
where \(\phi_{0}\) and \(\phi_{1}\) are set to \(\phi_{0}=\pi/2, \phi_{1}=0\) for the asymmetric dual-Bracewell mode and \(\phi_{0}=0, \phi_{1}=\pi\) for the symmetric dual-Bracewell mode. For the single-Bracewell mode, a corresponding matrix \(U\) can likewise be defined, subject to the condition that the beams have a relative phase orientation of \(\pi\) at the dark outputs. The transmission map for output \(m\) is then given by
\begin{equation} \label{Gl4}
T_m(\alpha_i,\beta_j,\lambda)
=
|W_m|^2
=
\left|
\sum_{k=0}^{N-1}
U_{m,k} V_k(\alpha_i,\beta_j,\lambda)
\right|^2.
\end{equation}
Substituting Eq.~\ref{Gl1} and Eq.~\ref{Gl3} into the definition of the transmission map (Eq.~\ref{Gl4}) yields the differential transmission map \(T_{\rm diff}=T_3-T_4\):
\begin{equation} \label{Gl6}
\begin{aligned}
T_{\rm diff}(\alpha_i,\beta_j,\lambda)
&\propto (1-b)\,
\mathrm{Re} \Bigg[
\Big( \sum_{k=0}^{1} e^{i \phi_0} (-1)^{k+1} r_k \,
e^{i \frac{2\pi}{\lambda} (x_k \alpha_i + y_k \beta_j)} \Big) \\
&\quad\cdot
\Big( \sum_{k=2}^{3} e^{i \phi_1} (-1)^{k+1} r_k \,
e^{i \frac{2\pi}{\lambda} (x_k \alpha_i + y_k \beta_j)} \Big)^*
\Bigg].
\end{aligned}
\end{equation}
This map provides the basis for the calculation of the differential photon flux rate \(P_{\mathrm{diff}}\) (see Eq.~\ref{Gl8}).


\bsp	
\label{lastpage}
\end{document}